\documentstyle [prl,aps,multicol,epsf]{revtex}
\begin{document}
\draft
\title{Brillouin Scattering Study of Propylene Carbonate:
An Evaluation of Phenomenological and Mode Coupling Analyses }
\author{Alexander Brodin$^{1}$, Martin Frank$^{1,2}$, Sabine Wiebel$^{2}$,
Guoqing Shen $^{1*}$, Joachim Wuttke$^{2}$, and H.Z. Cummins$^{1}$} 

\address{$^{1}$ Physics Department,
City College of the City University of New York, New York, NY 10031}
\address{$^{2}$ Physics Department E13,The Technical University of Munich,
D-85747 Garching, GERMANY}
\date{\today}
\maketitle
\begin{abstract}
Brillouin scattering spectra of the molecular glassformer propylene carbonate 
(PC) in the temperature range 140~K to 350~K were analyzed using both the 
phenomenological Cole-Davidson memory function and a hybrid memory
function consisting of the Cole-Davidson function plus a power-law term representing the critical decay part of the fast $\beta$ relaxation. 
The spectra were also analyzed using the extended two-correlator 
schematic MCT model recently employed by G\"{o}tze and Voigtmann to 
analyze depolarized light backscattering,
dielectric, and neutron-scattering spectra of PC [W.~G\"{o}tze and
Th.~Voigtman, Phys.\ Rev.\ E {\bf 61}, 4133 (2000)].  We assess the ability of 
the phenomenological and MCT fits, each with three
free fitting parameters, to simultaneously describe the spectra and give
reasonable values for the $\alpha$-relaxation time $\tau_{\alpha}$.

\end{abstract}

\begin{multicols}{2}

\section{INTRODUCTION}

Brillouin scattering spectroscopy has been one of the experimental techniques
widely applied to the study of glass-forming materials in the continuing
effort to achieve a complete understanding of the liquid-glass transition. 
The evolution of the Brillouin spectrum with temperature reflects the 
interaction of longitudinal sound waves ($q \sim 10^{5}$~cm$^{-1}$)
with structural relaxation.  Typically, as $T$ decreases from above the
melting temperature $T_{M}$ to below the glass-transition temperature
$T_{G}$, the Brillouin linewidth $\Delta \omega_{B}$ first increases, 
passes through a maximum, and then decreases again, while the Brillouin peak
position  $\omega_{B}$ shifts monotonically to higher frequencies, with the
increase occurring predominantly in the temperature range where 
$\Delta \omega_{B}$ is biggest. 

In the earliest experiments, only the temperature dependence of the Brillouin
shift and linewidth were studied.  Subsequently, the full spectrum was analyzed 
and compared with theoretical predictions derived primarily
from generalized hydrodynamics models.  Either the Mountain
 formalism [Mountain~1968] or a generalized damped harmonic 
oscillator model were used in which the 
memory function, i.e.\ the frequency-dependent part of the longitudinal
viscosity, represents the structural relaxation dynamics.  Generally, 
acceptable fits could be obtained using an exponential or 
stretched-exponential empirical memory function $m(t)$ 
(or a Cole-Davidson memory function $m[\omega]$), 
with the relaxation time $\tau(T)$ 
treated as an adjustable fitting parameter.  However, the $\tau(T)$ values 
determined from such fits were usually found to increase much more 
slowly with decreasing $T$ at low temperatures than the values 
determined with other techniques such as dielectric spectroscopy or ultrasonics.

The origin of this apparent disagreement can be attributed to the use of
these ``$\alpha$-relaxation only'' models for $m(t)$ which ignore the fast
part of the relaxation process preceding the final $\alpha$-relaxation. 
The need for an ``additional fast contribution'' to $m(\omega)$ was first
 noted by Loheider {\it et al} [Loheider~1990] (also see [Cummins~1994]).
Several empirical approaches to extending $m(\omega)$ or $m(t)$ to include
the missing fast $\beta$ relaxation have been described in the literature,
including: (1)~adding a second Debye, Cole-Cole, or Cole-Davidson function 
to the Cole-Davidson $\alpha$-relaxation term [Soltwisch~1998]
[Monaco~2000, Monaco~2001];
(2)~extracting a memory function including both $\alpha$ and $\beta$
relaxation from depolarized backscattering
spectra and using it for fitting the Brillouin spectra [Cummins~1994] 
(a procedure that succeeded for CKN but not for other materials); 
(3) taking $m(t)$ as the sum of a stretched exponential plus a damped
high-frequency oscillatory term [Lebon~1997]; and (4)~adding a term
$B \omega^{a}$ to the Cole-Davidson function to approximate
the critical-decay component of the fast $\beta$-relaxation predicted
by the mode coupling theory (MCT) [Du~1994, 1996, Cummins~1994,
Pick~2000].  This additive ``hybrid model'' approach, initially suggested 
by G\"{o}tze [Gotze~1993], has recently been shown to provide excellent
fits to depolarized backscattering spectra of toluene [Wiedersich~2000] and has
also been used in the analysis of depolarized Brillouin spectra of metatoluidine 
[Pick~2000].  

With the additional fast relaxation included through any of these
approaches, the $\alpha$-relaxation time $\tau_{\alpha}$ is found to
increase much more rapidly with decreasing temperature than it does
with $\alpha$-only models, consistent with other experimental 
measurements.  However, while this approach can resolve the 
$\tau(T)$ disagreement, it usually requires additional fitting parameters
and does not explain the origin of the structural relaxation dynamics that
it describes.  Furthermore, various authors have chosen quite different
empirical fitting functions and have usually obtained good fits to
Brillouin spectra, suggesting that this procedure is not able to distinguish 
effectively between different models.

Another approach to the analysis of Brillouin scattering spectra utilizes
the mode coupling theory (MCT) in which the slowing down of structural
relaxation with decreasing $T$ arises naturally from nonlinear interactions
among density fluctuations.  A major advantage of this approach is that both
the $\alpha$ and $\beta$ regions of the relaxation dynamics occur spontaneously 
in the solutions of the MCT equations.  However, while MCT can provide a
quantitative microscopic description of liquid dynamics in principle,
it requires the intermolecular potential [or the static structure factors
$S(q)$] as input, and this information is not  generally available for 
most real glassforming materials.   Therefore, MCT analyses are usually
 restricted to comparison with asymptotic
predictions in which the detailed microscopic interaction information is
absorbed into a single parameter, the exponent parameter $\lambda$.
Alternatively, a simplified ``schematic MCT'' can be used in which the 
full set of density correlators is replaced by a small number (usually
one or two), and the coupling constants are treated as adjustable fitting
parameters.  This schematic approach captures the basic mechanism underlying
the MCT more fully than the asymptotic analyses, but sacrifices the 
material-specific microscopic interpretation of the parameters.

A convenient schematic MCT model, initially introduced by Sj\"{o}gren 
[Sj\"{o}gren~1986], uses two correlators: $\phi(t)$ for the ``system'' and 
$\phi_{s}(t)$ for the ``probe'', i.e.\ the variable being measured in a particular
experiment.  In this Sj\"{o}gren model, the interesting critical behavior
occurs in $\phi(t)$ while the temperature dependence of $\phi_{s}(t)$
results from its being coupled to $\phi(t)$.  This model was successfully
used to analyze depolarized light-scattering spectra of glycerol
[Franosch~1997] and OTP [Singh~1998].  Ruffl\'{e} {\it et al} 
[Ruffl\'{e}~1999] showed that Brillouin scattering spectra of 
NaLi(PO$_{3}$)$_{2}$ could also be analyzed with the Sj\"{o}gren model, 
using parameters for the system correlator $\phi(t)$ determined from
inelastic neutron scattering spectra.

Recently, G\"{o}tze and Voigtmann have extended the Sj\"{o}gren 
model by including hopping parameters $\Delta$ and $\Delta_{s}$
in the equations of 
motion for both $\phi(t)$ and $\phi_{s}(t)$, which allows the analysis
to be extended to low temperatures where relaxation is dominated by 
activated hopping processes. 
They analyzed dielectric, depolarized light scattering,
and neutron scattering data for propylene carbonate (PC), using a single set
of system parameters for $\phi(t)$, optimized simultaneously for all the 
experiments [Gotze~2000].  This result suggests the possibility of analyzing
PC Brillouin scattering spectra using the system parameters already determined
by G\"{o}tze and Voigtmann.  Moreover, if the probe hopping parameter
$\Delta_{s}$, frequency $\Omega_{s}$, and damping $\nu_{s}$ are taken as
fixed by the depolarized light scattering fits, then the analysis will include only
one adjustable MCT fitting parameter, the coupling constant $V_{s}$, 
which determines the strength of coupling between $\phi(t)$ and $\phi_{s}(t)$.
The fits should then {\em simultaneously} describe the Brillouin spectra
and also determine the value
of the structural relaxation time $\tau_{\alpha}$.  Such an analysis could 
provide a stringent test of the utility of MCT for the analysis of Brillouin 
scattering spectra.  A test of this procedure was the principal motivation for 
the present Brillouin scattering study of PC.

We describe the experiments in Sec.~II.  We analyze the data in
Secs.~III and IV using two empirical memory-function models, first a conventional
$\alpha$-relaxation-only Cole-Davidson~(C.D.) function and then the
hybrid model.  (A preliminary version of the C.D.\ and hybrid fits was 
presented by Frank [Frank~2000].)  In Sec.~IV we also include a 
discussion of the role of thermal diffusion, followed by a second
hybrid model analysis in which 
the C.D. stretching parameter $\beta$ is constrained to be equal to the
MCT critical exponent $b$.  In Sec.~V we present our analysis of the 
spectra with the two-correlator extended schematic MCT approach.  
In Sec.~VI, we provide discussion and conclusions including a brief 
evaluation of the ability of these and related approaches to extract 
meaningful information about structural relaxation dynamics from the
analysis of Brillouin scattering spectra.

\section{Experiment}

\subsection{Material}

Propylene carbonate (PC; $T_{G} = 160$ K) is a molecular glassforming
material that has been extensively studied with a wide range of experimental
techniques including Brillouin scattering [Elmroth~1992; Du~1994].
In Table~I we list the principal properties of PC that are
relevant for the present study.

A recent dielectric spectroscopy study of PC by Schneider {\it et al}
[Schneider~1999] determined the complex dielectric constant
$\epsilon(\omega)$ over a frequency
range of 18 decades for temperatures from 153~K to 293~K.  
Cole-Davidson fits to the $\alpha$ peaks in the $\epsilon''(\omega)$
spectra produced the $\tau_{\alpha}^{\epsilon}(T)$ values shown in 
Fig.~1
by the open circles.  Schneider {\it et al} carried out fits to these
values with several empirical fitting functions.  Their extended free
volume fit is shown in Fig.~1 by the solid line.

Du {\it et al} [Du~1994] analyzed the $\alpha$ peaks in $\chi ''(\omega)$
spectra, obtained from their depolarized PC backscattering spectra, with a
Kohlrausch function. The resulting $\tau_{\alpha}^{DLS}$ values, shown
in Fig.~1 by squares, have temperature dependence similar to the dielectric
$\tau_{\alpha}^{\epsilon}(T)$ although they are  approximately three times smaller.

%
\vbox{
\vspace{0.2 in}
\hbox{
\hspace{0in}
\epsfxsize 3.0in \epsfbox{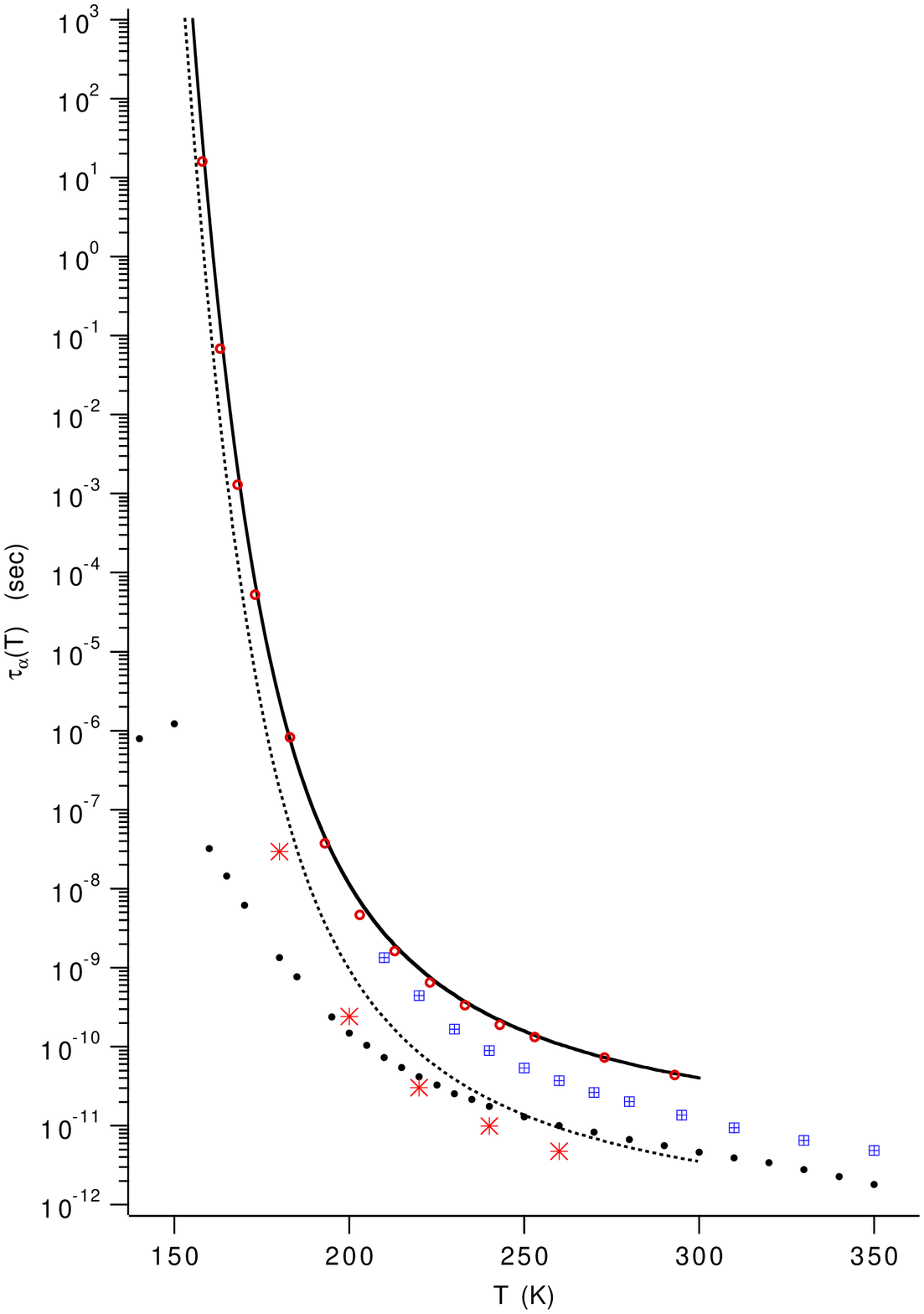}
}
\vspace{0in}
}
\refstepcounter{figure}
\parbox[b]{3.3in}{\baselineskip=12pt FIG.~\thefigure.
(TAU). Propylene carbonate (PC) 
relaxation time $\tau_{\alpha}(T)$.  Open circles and solid line:
dielectric measurements and extended free-volume fits [Schneider~1999].
Squares: depolarized light scattering spectra [Du~1994].
Solid circles: Brillouin scattering $\tau^{CD}_{\alpha}(T)$
from Cole-Davidson fits (Sec.~III).  Broken line: dielectric 
free volume fit multiplied by 0.085 to match Brillouin $\tau_{\alpha}(T)$
in the region of the maximum Brillouin linewidth.  $\ast$: 
$\tau_{\alpha}^{MCT}$ obtained from the extended schematic MCT fits.
\vspace{0.10in}
}
\label{1}
\vspace{0.1in}

The analysis of our Brillouin scattering spectra using a
Cole-Davidson memory function, described in the next section, resulted
in the $\tau_{\alpha}^{CD}$ values shown
by the solid circles in Fig.~1.  As mentioned in the Introduction,
at low temperature these values do not increase as rapidly with decreasing
temperature as the dielectric and backscattering results, 
suggesting that the Cole-Davidson function provides an
incomplete representation of structural relaxation.  
(At $T = T_{G}$, $\tau_{\alpha}^{CD}$ has only reached
$\sim 30$~ns, while typically $\tau_{\alpha}(T_{G})$ is $\sim 100$~s.)

In our $90^{\circ}$ PC Brillouin scattering experiment, the maximum
Brillouin linewidth occurs at $T \sim 230$~K.
At that temperature the peak of
the $\alpha$ relaxation coincides with the Brillouin line, so 
$\omega_{B}\tau_{\alpha}^{LA} \approx 1$ which gives
$\tau_{\alpha}^{LA}$ (230)
$\sim 3.2 \times 10^{-11}$ sec, a value $\sim 12$ times smaller than the
$\tau_{\alpha}^{\epsilon}$ of Schneider {\it et al}.
The difference presumably
arises from the fact that $\tau_{\alpha}^{LA}$ characterizes
longitudinal displacement dynamics while $\tau_{\alpha}^{\varepsilon}$
characterizes orientational dynamics.
In Fig.~1, the lower line
is the free volume fit of Schneider {\it et al} divided by a factor
of 11.75 to match $\tau_{\alpha}^{LA}$ in this region:

\begin{equation}
\log_{10} ( \tau_{\alpha}^{LA}) =
 -A + B/  \{ T-T_{0} + [(T-T_{0})^{2} + CT]^{1/2} \}
\end{equation}
with $A=12.56$, $B=309$, $C=4.82$, and $T_{0} = 162$~K 
($\tau_{\alpha}^{LA}$ in seconds).   In constructing memory functions
for data analysis beyond the Cole-Davidson model, we will use Eq.~(1)
to fix the $\alpha$-relaxation time of the longitudinal acoustic mode
$\tau_{\alpha}^{LA}$.

\subsection{Sample Preparation}

Propylene carbonate (anhydrous, 99.7~\%) was purchased from 
Sigma-Aldrich.  PC was loaded in a glove box under dry nitrogen
atmosphere into a distillation flask fitted with a stopcock.
The flask was then transferred to a vacuum distillation system
and distilled at $\sim 110$~C into glass sample cells that were
flame sealed under vacuum.  A sample cell was installed in an Oxford
LN2 coldfinger cryostat with an ITC-4 temperature controller for
the Brillouin scattering experiments. 

\subsection{Brillouin scattering experiments}

Spectra were collected in both 90$^{\circ}$~VV geometry and
174$^{\circ}$~VH near-backscattering geometry with a Sandercock 
6-pass tandem Fabry-Perot interferometer at temperatures from 140~K to
350~K in 5~K or 10~K steps.  The 514.5~nm single-mode argon laser 
power at the sample was typically 170~mW.  The interferometer finesse 
was $\sim~90$; its contrast was better than $10^{7}$.  Complete sets of 
spectra were collected with mirror separations $d=2$, 5, and 10~mm for
VV and $d=2$~mm and 10~mm for VH.  Spectra were accumulated 
during 1 to 2 hour runs.  For the $d=10$~mm separation spectra two 
2-hour runs were collected and averaged to improve the signal-to-noise
ratio.  In Fig.~2 we plot the Stokes side of the three VV spectra (a)
and the two VH spectra (b) for $T=220$~K. 
(The dark counts have already been subtracted.)

In order to test for possible experimental artifacts, $I_{ISO}(\omega)$
PC Brillouin spectra at several temperatures were collected at both the
Technical University of Munich and at the City College of New York
with different samples, different tandem Fabry-Perot spectrometers, and
different experimental procedures.  The spectra obtained were in very good
agreement.  The spectra shown in this work were all collected in New York.
%
%
\vbox{
\vspace{0.2 in}
\hbox{
\hspace{0in}
\epsfxsize 3.0in \epsfbox{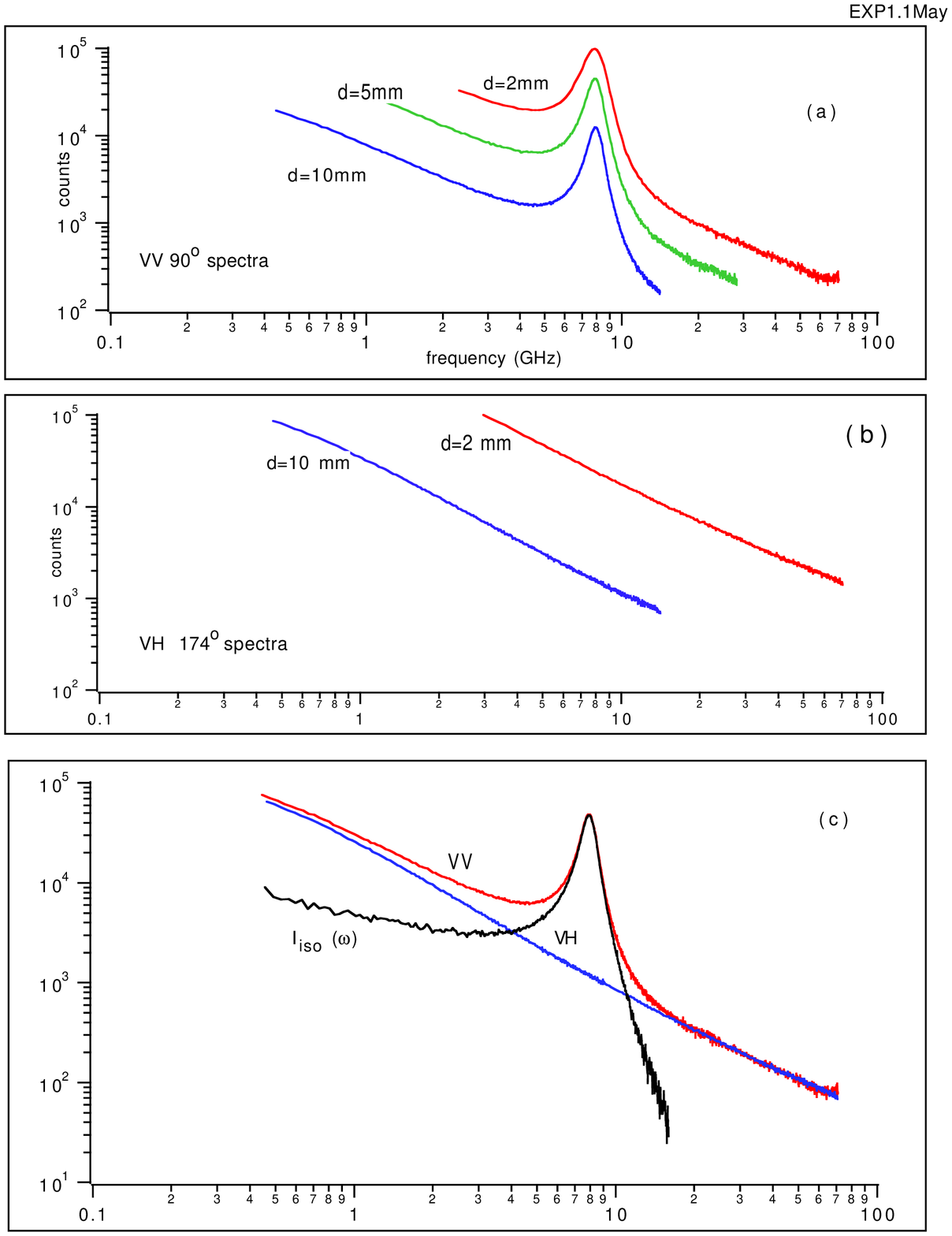}
}
\vspace{0.1in}
}
\refstepcounter{figure}
\parbox[b]{3.3in}{\baselineskip=12pt FIG.~\thefigure.
(EXP1). (a)~Stokes sides of 220~K PC ~$90^{\circ}$ VV Brillouin spectra 
with $d = 2,5$,and 10~mm. (b)~$174^{\circ}$ VH spectra with $d = 2$~mm
and 10~mm.  (c)~The composite 90$^{\circ}$ VV spectrum obtained by shifting the
2~mm and 5~mm spectra of (a) to optimize overlap with the 10~mm spectrum,
the composite VH spectrum obtained similarly and then shifted to optimize
overlap with the VV spectrum for frequencies above 30~GHz, and the
difference spectrum $I_{ISO}(\omega)$ obtained by subtraction.  
The background from dark counts has already been subtracted in the 
spectra of (a) and (b).}
\vspace{0.10in}

\label{1}
\vspace{0.1in}

\subsection{Data reduction: extraction of the density fluctuation spectra 
$I_{ISO}(\omega)$}

The VV polarized Brillouin spectra contain contributions from 
density fluctuations [$I_{ISO}(\omega)$] and also from
orientational dynamics and collision-induced scattering
[$I_{ANI}(\omega)$]. Depolarized (VH) backscattering spectra
are due to $I_{ANI}(\omega)$ only, and can therefore be used
to remove the $I_{ANI}(\omega)$ contribution from the VV spectra.

Fioretto {\it et al} [Fioretto~1999] [Monaco~2001] have carried out the
subtraction by combining spectra obtained with different mirror 
separations to obtain broadband VV and VH spectra, and scaling the VH
spectra to overlap the VV spectra for frequencies above $\sim 40$~GHz
where $I_{ISO}(\omega)$ should be negligible.  Subtracting 
the scaled $I_{VH}(\omega)$ from the $I_{VV}(\omega)$ then
provides $I_{ISO}(\omega)$.  We have also followed this procedure.

The Stokes sides of the $d=2$ and 10~mm 220~K VV spectra, shown in 
Fig.~2(a), were rescaled vertically to optimize 
overlap with the $d=5$~mm~GHz spectrum, producing the 
composite $90^{\circ}$ $I_{VV}(\omega)$ spectrum shown in Fig.~2(c).
The $\theta = 184^{\circ}$ $I_{VH}(\omega)$ spectra were similarly
combined, and the composite $I_{VH}$ spectrum was again scaled vertically
to match the $I_{VV}$ spectrum in the 30~-~50~GHz region. 
Finally, the scaled $I_{VH}$ spectrum was subtracted from the
$I_{VV}$ spectrum to produce the difference spectrum 
$I_{ISO}(\omega)$ as shown in Fig.~2(c).

The full set of $I_{ISO}(\omega)$ spectra for temperatures from 140~K
to 350~K obtained with this procedure is shown in 
Fig.~3.
For each set of VV spectra, we also recorded an instrument profile 
spectrum with $d=10$~mm.  The width of the instrumental profile was
typically FWHH = 0.17~GHz.  To reduce additional Brillouin peak
broadening due to the finite input aperture (which leads to collecting light
scattered at slightly different angles), the input aperture was reduced
sufficiently to decrease the additional broadening at the Brillouin 
frequency to less than the instrumental width.  This additional geometrical
broadening (typically $\lesssim 0.13$~GHz) was then calculated for every
temperature from the known collection geometry and convoluted into the
instrument function.  In the fitting procedures described in this paper, 
the instrument function so obtained was convoluted with the theoretical
spectrum for comparison with the $I_{ISO}(\omega)$ spectrum.
The fits were carried out with a conventional nonlinear least-squares
fitting program (NLLSQ).  

We first fit each Brillouin peak in the $d=10$~mm $VV$ spectra to the 
damped harmonic oscillator function
%
%
\vbox{
\vspace{0.2 in}
\hbox{
\hspace{0in}
\epsfxsize 3.0in \epsfbox{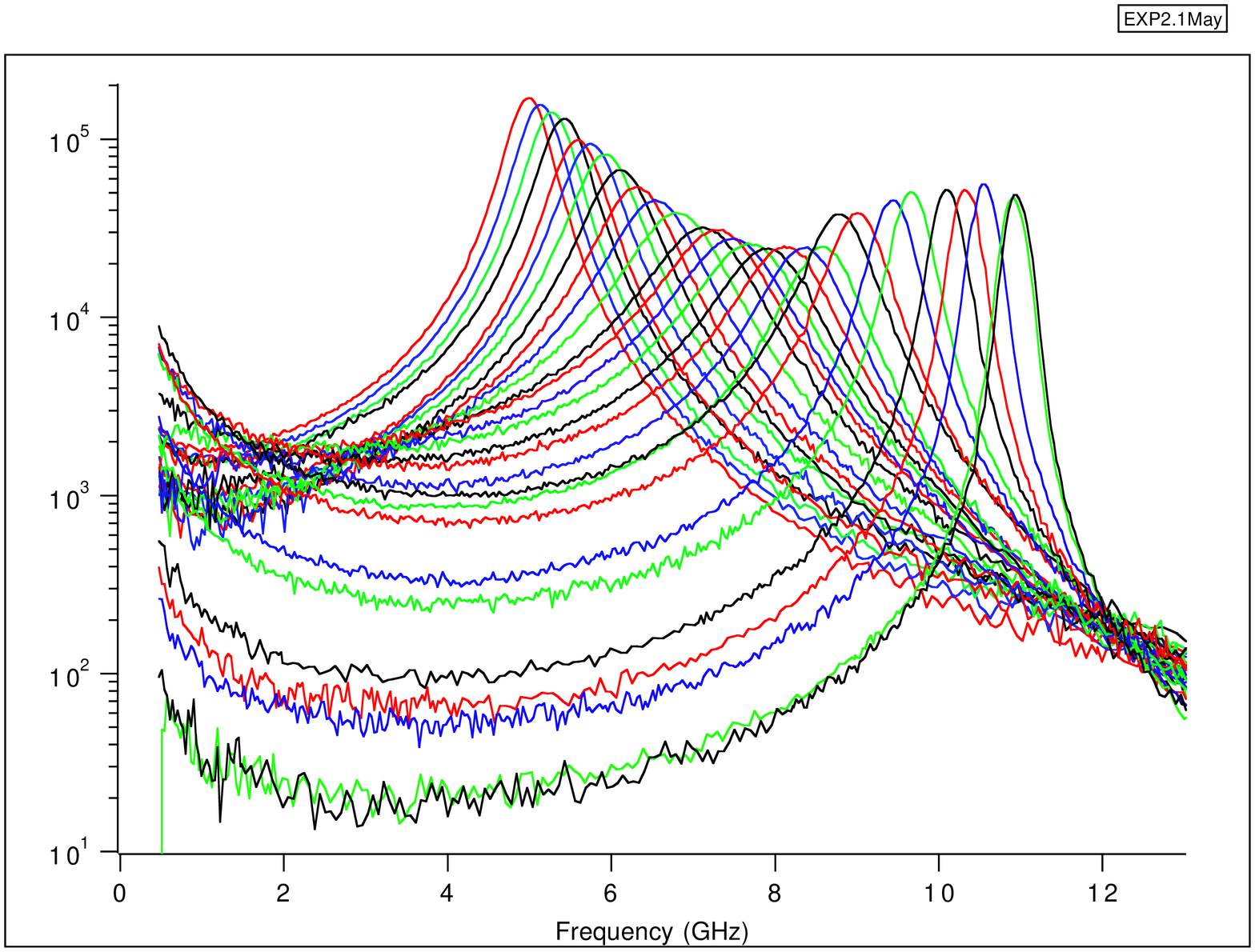}
}
\vspace{0.1in}
}
\refstepcounter{figure}
\parbox[b]{3.3in}{\baselineskip=12pt FIG.~\thefigure.
(EXP2).  Complete set of composite 
$I_{ISO}(\omega)$ spectra for 28
temperatures from 140~K to 350~K obtained by the procedure shown in 
Fig.~2.  $T = 140$, 150, 160, 165, 170, 180, 185, 195, 200, 205,
210, 215, 220, 225, 230, 235, 240, 250, 260, 270, 280, 290, 300, 310,
320, 330, 340, and 350~K, from right to left.}
\vspace{0.10in}

\label{1}
\vspace{0.1in}

\vspace {0.1in}
\begin{eqnarray}
I(\omega) = \frac{I_{0}}{\omega} {\cal{I}}m [\omega^{2}-\omega_{B}^{2}
-i \omega \gamma_{B} ]^{-1} \nonumber\\
=\frac{I_{0}\gamma_{B}}{ [\omega^{2}-\omega_{B}^{2}]^{2} + [\omega \gamma_{B}]^{2}} \ 
\end{eqnarray}
including convolution with the instrument function.  In 
Fig.~4~ the
resulting values of $\omega_{B}(a)$ and $\gamma_{B}(b)$ are shown
by the points.  For a simple liquid, both $\omega_{B}$ and
$\gamma_{B}$ would have simple monotonic temperature dependence.
The observed temperature dependence of $\omega_{B}$ and $\gamma_{B}$,
as noted in the introduction, is the signature of structural relaxation 
dynamics as seen in Brillouin scattering. The maximum in $\gamma_{B}$
at $T \sim$~230~K indicates that the $\alpha$ peak moves through the
Brillouin line at that temperature.

\section{Data Analysis 1: The Cole-Davidson Model}

\subsection{The memory function approach}

The isotropic Brillouin spectrum $I_{ISO}(\omega)$ is usually
attributed to density fluctuations $\rho_{q}(t)$ where the scattering
vector $q = (2 \pi  n / \lambda) \sin (\theta/2)$.  The spectrum consists
of a triplet: the central thermal diffusion mode (which is generally 
located at frequencies too low to observe in these experiments and will
be ignored in most of the fits) and the Brillouin doublet.  The spectrum is
conventionally described by

\begin{equation}
I(\omega) = \frac{I_{0}[\gamma_{0} + m''(\omega)]}{[\omega^{2}-
\omega_{0}^{2} + \omega m'(\omega)]^{2} + [\omega \gamma_{0} +
\omega m''(\omega)]^{2}} \ 
\end{equation}
[The particular Laplace transform convention used to derive Eq.~(3) is
discussed in the Appendix.]

Equation (3) is very general; $m(\omega) = m'(\omega)+i m ''(\omega)$
is the Laplace transform of the memory function $m(t)$ which can be
constructed to include structural relaxation, thermal diffusion, and
translation-orientation coupling for liquids of anisotropic molecules
[Chung~1971; Boon~1980; Scopigno~2000].  Usually, however, $m(t)$
represents the relaxing longitudinal viscosity, first introduced
by Mountain~[1968] to describe the damping of sound waves by
interaction with internal degrees of freedom of the molecules, and
later utilized to represent the interaction of sound waves with
structural relaxation.  (Both mechanisms may
contribute, however, as discussed below.)  In the long-wavelength limit,
$\omega_{0} = C_{0}q$ where $C_{0}$ is the limiting low-frequency
adiabatic sound velocity, and $\gamma_{0}$ is the ``regular'' sound
attenuation coefficient.  Note that Eq.~(3) is the equation for
a damped harmonic oscillator with a frequency-dependent
damping function $\gamma(\omega) = \gamma_{0} + m(\omega)$.

Ideally, the $I_{ISO}(\omega)$ spectra should be analyzed to yield
$m(\omega)$ directly.  But in practice this is not possible because
the spectra only cover a range of two decades at most, while
$m(\omega)$ extends over many decades. Also, the spectra are
dominated by the Brillouin components [which would be the complete spectrum
of Eq.~(3) if $m(\omega)=0$], and the modification of $I(\omega)$
produced by inclusion of $m(\omega)$ is not very sensitive to the
detailed form of $m(\omega)$.  Therefore, data analysis almost always
proceeds by selecting a parameterized model for $m(\omega)$ and varying
the parameters at each temperature to optimize the fits.

The primary requirements on $m(\omega)$ are that $m'(\omega_{B})$,
the real part of $m(\omega)$ at the frequency of the Brillouin peak, 
must be adjusted to shift the Brillouin peak from $\omega_{0}$ to
$\omega_{B}$; and $m''(\omega_{B})$, the imaginary part of $m(\omega)$
at the frequency of the Brillouin peak, must be adjusted so that
$\gamma_{0} + m''(\omega_{B})$ will give the correct Brillouin linewidth.
If these two conditions are simultaneously satisfied, then the particular form 
chosen for $m(\omega)$ will only influence the details of the lineshape.
Therefore the essential requirement is that $m''(\omega_{B})/m'(\omega_{B})$
must have the ``correct'' value.  Nevertheless, this is not a trivial requirement
since $m'(\omega)$ and $m''(\omega)$ are connected by Kramers-Kronig relations.

The prototype memory function is the exponential $e^{-t/\tau}$,
multiplied by a coupling constant $\Delta^{2}$.  (It is usually called Debye
relaxation although it was first used by Maxwell in his theory of 
viscoelasticity.)  The simplest memory function $m(t)$ for structural 
relaxation that fits various experimental data in the frequency region 
of $\alpha$ relaxation is the stretched-exponential (KWW) function 
$e^{-(t/\tau)^{\beta_{K}}}$.  An approximation to the Fourier transform
of this $m(t)$ is provided by the Cole-Davidson~(C.D.) function

\begin{equation}
m(\omega) = (\Delta^{2}/\omega) [(1-i\omega \tau)^{-\beta} -1] \ 
\end{equation}
which has frequently been used, together with Eq.~(3), to analyze
Brillouin-scattering spectra.  In practice, the fitting procedure 
adjusts the coupling constant $\Delta^{2}$ so that the real part
of $m(\omega)$ produces the correct $\omega_{B}$ for the Brillouin peak, while 
$\tau$ is adjusted to give the correct Brillouin linewidth.

There are several problems that should be noted in carrying out this analysis,
as well as analyses with more elaborate memory functions.

(1)  Many glassforming materials are molecular liquids, and internal
vibrational modes may also contribute to $m(\omega)$, as originally
proposed by Mountain [1968].  Recently, Monaco {\it et al}
([Monaco~2000]) have carried out a Brillouin scattering study of OTP from 
which they conclude, for OTP, that the fast part of the relaxation process is
{\em entirely} due to intramolecular vibrational modes.  The assignment
results from the presence of additional low frequency structure 
(Mountain mode) in VV, but not VH, spectra, in both the glass and the
crystal.  This conclusion has recently found additional support from a 
molecular dynamics study of OTP [Mossa~2001].  There is no evidence, 
however, that it applies to other materials such as PC, and it will not be 
considered here.

(2)  The identification of $I_{ISO}(\omega)$ with density fluctuation rests
on the assumption that scattering due to orientational fluctuations,
which is typical for molecular glassforming materials, is eliminated by 
the subtraction of $I_{ani}(\omega)$ from $I_{VV}(\omega)$ described in
the previous section.  However, two consequences of anisotropy may remain
after subtraction.  First, since longitudinal current involves both compression
and shear, rotation-translation coupling modifies the longitudinal current 
correlator [which determines the spectrum of $\rho_{q}(t)$] as shown in
Eq.~(A.14) of Ref.~[Dreyfus~1999].  Second, as noted recently by Latz 
and Letz [Latz~2000], because of this coupling the orientational dynamics
also reflect the longitudinal acoustic mode, and this part of the orientational
dynamics is not removed by subtraction.  Therefore, both density fluctuations
and orientational dynamics may contribute to the $I_{ISO}(\omega)$ spectra.

While the extent of these corrections due to molecular anisotropy is not yet
clear, preliminary simulations, carried out in collaboration with R.M.~Pick,
indicate that they are relatively minor.  We will therefore use the
conventional memory function formalism in this paper, recognizing that the
values of the parameters used to describe the frequency-dependent longitudinal
viscosity may reflect some aspects of translation-rotation coupling and 
molecular anisotropy and may also include some contribution from 
intramolecular vibrational dynamics.

\subsection{Cole-Davidson Fits}

We first analyzed the $I_{ISO}(\omega)$ spectra of Fig.~3~ using 
Eq.~(3) with the C.D.\ memory function $m(\omega)$ of Eq.~(4).
The fit results, for a subset of the spectra, 
are shown in Fig.~5 [note (IIIB.1)].
As often found in the past, excellent fits can be obtained if all the parameters
are kept free.  However, it is preferable to fix as many parameters as possible
using independently determined values, since the parameters are rather strongly
coupled.  (This coupling can also lead to unstable parameter values in the 
fit results.)  Here we consider each of the relevant parameters separately.

(1) $\beta$.  The fits are relatively insensitive to this parameter.  We took
$\beta = 0.68$ (corresponding to $\beta_{K}= 0.77$) as found in the analysis
of depolarized backscattering spectra by Du {\it et al} [Du~1994].

(2) $\omega_{0}(T)$.  Since $\omega_{0} = C_{0}q$ where $C_{0}$ 
is the limiting low-frequency adiabatic sound velocity and $q$ is the scattering wavevector $(4 \pi n / \lambda) \sin (\theta/2)$, $\omega_{0}(T)$ can be fixed
if $C_{0}(T)$ and the refractive index $n(T)$ are known.
$C_{0}(T)$ was previously measured by ultrasonic experiments at 5~MHz
and 15~MHz by Du {\it et al} [Du 1994] which gave $C_{0}(T) = 2.507 
\times 10^{5} -361.2$ $T$~cm/sec.
%
\vbox{
\vspace{0.2 in}
\hbox{
\hspace{0in}
\epsfxsize 3.2in \epsfbox{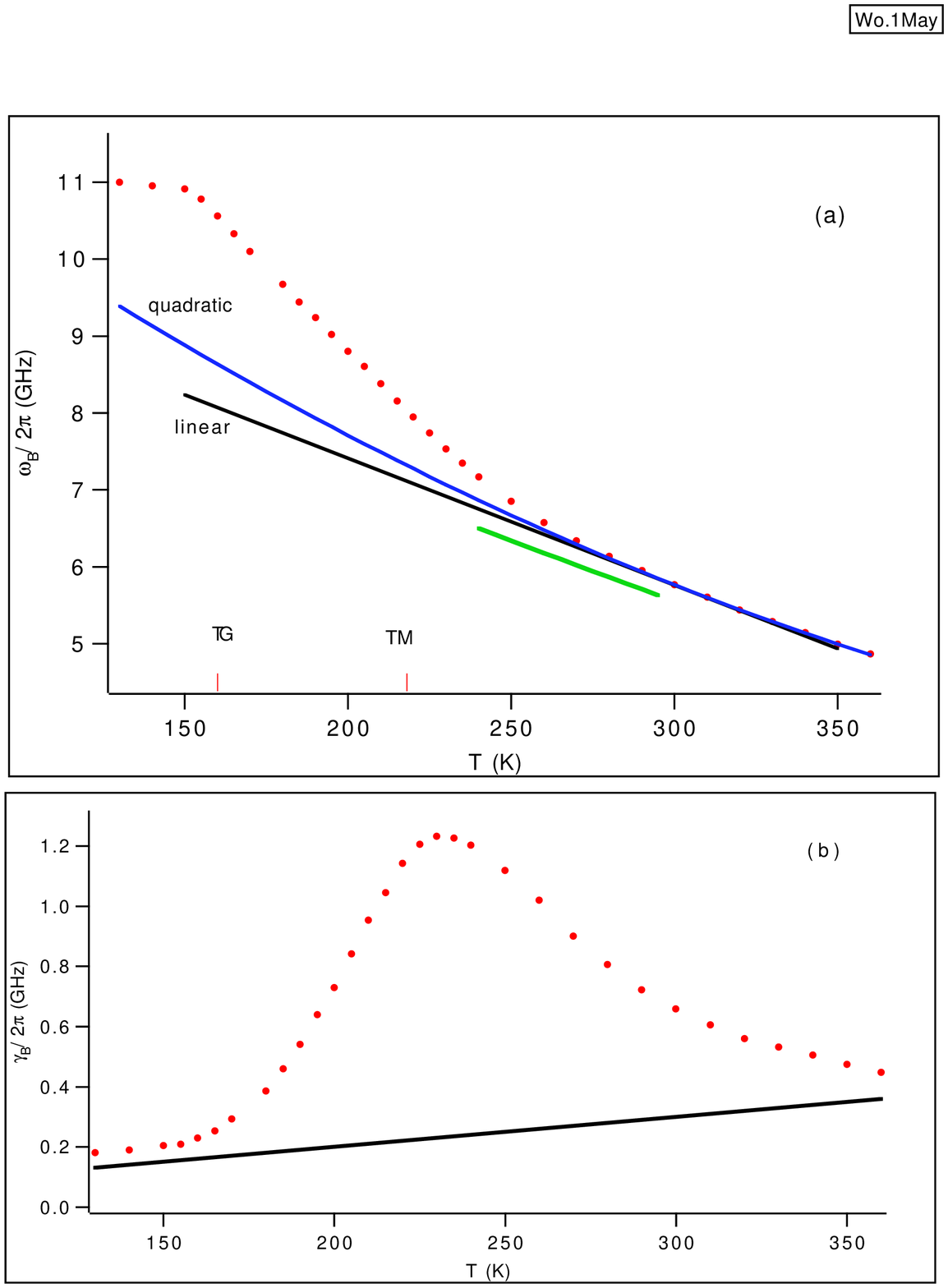}
}
\vspace{0.1in}
}
\refstepcounter{figure}
\parbox[b]{3.3in}{\baselineskip=12pt FIG.~\thefigure.
(Wo).  Fits of the $d=10$~mm $90^{\circ}$
VV Brillouin peaks to Eq.~(2) including convolution with the
instrument function. (a)~$\omega_{B}/2\pi$(GHz) (circles).
Short solid line: $\omega_{0}$ from ultrasonic and refractive 
index data; long solid line: $\omega_{0}$ increased by 3.5\% 
to match $\omega_{B}$ in the high-temperature region.  Solid curved
line: quadratic fit $\omega_{0}(T) = 13.211 - 3.291 \times 10^{-2}$~T 
$+ 2.6973 \times 10^{-5}$ T$^{2}$. (b) $\gamma_{B}/2\pi$~(GHz)
(circles).  Solid line: $\gamma_{0} = T/1000$.}

\vspace{0.10in}

\label{1}
\vspace{0.1in}

  However, the ultrasonic data only cover
the range 240 to 293~K.  At lower temperatures, there is significant dispersion
in the ultrasonic sound speed, while higher temperature measurements are 
technically difficult.  The refractive index $n(T) = 1.5314 -3.752 \times 
10^{-4}$~$T$ (see Table~I) corresponds to the sodium $D$ line 
and may be different at 514.5~nm.  The $\omega_{0}(T)$ values 
implied by these measurements are shown in the upper panel of 
Fig.~4, by the short solid line.  A linear
extrapolation of these values, shifted up by 3.5 \%, is indicated by the long
solid line labeled ``linear'': $\omega_{0}(T)/2\pi = (10.71 - 0.0165 \times
T$)~GHz.  The shift, which is within the experimental errors of the $C_{0}(T)$
and $n(T)$ data, was chosen so that at high temperatures the Brillouin peak 
frequency $\omega_{B}$ is equal to $\omega_{0}$, as expected.  Note, however,
that the high-$T$ $\omega_{B}$ values suggest some small upward curvature
away from the linear extrapolation above 320~K.

Initially, fits were carried out with this linear-$T$ estimate of 
$\omega_{0}(T)$.  However, the fits were not all satisfactory, 
in part because of the high-$T$ curvature in $\omega_{B}$ seen in
Fig.~4.  For materials where $C_{0}(T)$ measurements extend 
over larger  temperature ranges, there can be significant upward 
curvature to $C_{0}(T)$.  We therefore tried fitting $\omega_{0}(T)$
to the six highest temperature $\omega_{B}(T)$ values plus one 
$\omega_{0}(T)$ value obtained by a free C.D.\ fit at 230~K where 
the Brillouin linewidth is largest.  The result was the quadratic 
$\omega_{0}(T)$ function shown by the upper solid line in 
Fig.~4~ labeled ``quadratic'':
\begin{equation}
\omega_{0}(T) = 13.211 - 3.291 \times 10^{-2} \ T + 2.6973
\times 10^{-5} \ T^{2} \  
\end{equation}
which was used in all subsequent fits.

(3)  $\gamma_{0}(T)$.  This term is conventionally included to represent
``regular'' damping of the sound waves by anharmonic processes not 
related to structural relaxation.  It can also represent a first frequency-independent 
approximation for the fast part of the structural relaxation not included in
$\alpha$-relaxation-only models such as the C.D.\ function.  Usually, 
$\gamma_{0}$ is taken as a temperature-independent constant, fixed 
from the Brillouin linewidth at low temperatures.

However, since anharmonic damping processes often tend to become 
stronger with increasing temperature, $\gamma_{0}$ may increase with
$T$.  Also, at temperatures well above the temperature of the maximum
Brillouin linewidth, the linewidth due to $\gamma_{0}$ plus structural
relaxation is approximately given by 
$\Delta \omega_{B} \approx \gamma_{0} + \Delta^{2} \tau \beta$.  
Fits with constant $\gamma_{0}$ often give $\Delta^{2}$
values that increase at high temperatures, a result that appears 
unphysical. This apparent increase in $\Delta^{2}$ can be avoided by
allowing $\gamma_{0}$ to increase with increasing $T$.  To avoid 
introducing additional fitting parameters we arbitrarily assumed 
$\gamma_{0}$ to be a linear function of $T$, and took 
$\gamma_{0}(T) = T/1000$ as shown by the solid line in the 
lower panel of Fig.~4.

(4)  $\Delta^{2}(T)$.  Since at high temperatures 
$\omega_{B} = \omega_{0}$ while 
$\Delta \omega_{B} = \gamma_{0} + \Delta^{2}\tau \beta$, 
the effective fitting parameter for $\Delta \omega_{B}$ is the product
$\Delta^{2}\tau$.  So if $\Delta^{2}$ and $\tau$ are both free parameters, 
the fits tend to be unstable.  We therefore carried out the fits for temperatures
up to 235~K with the three free fitting parameters: $\Delta^{2}$, $\tau$, and the
scale factor $I_{0}$.  For $T > 235$~K, $\Delta^{2}$ was fixed 
at its value at $\sim 235$~K, and the fits were carried out with only
$\tau$ and $I_{0}$ free.  
The parameters $\tau(T)$ and $\Delta^{2}(T)$
obtained from these fits are shown in Table~II along with the reduced 
$\chi^{2}$ value for each fit.
The C.D.\ fits, shown in Fig.~5, are satisfactory although better
fits could be obtained by keeping more parameters free, especially
$\omega_{0}$.
%
\vbox{
\vspace{0.2 in}
\hbox{
\hspace{0in}
\epsfxsize 3.2in \epsfbox{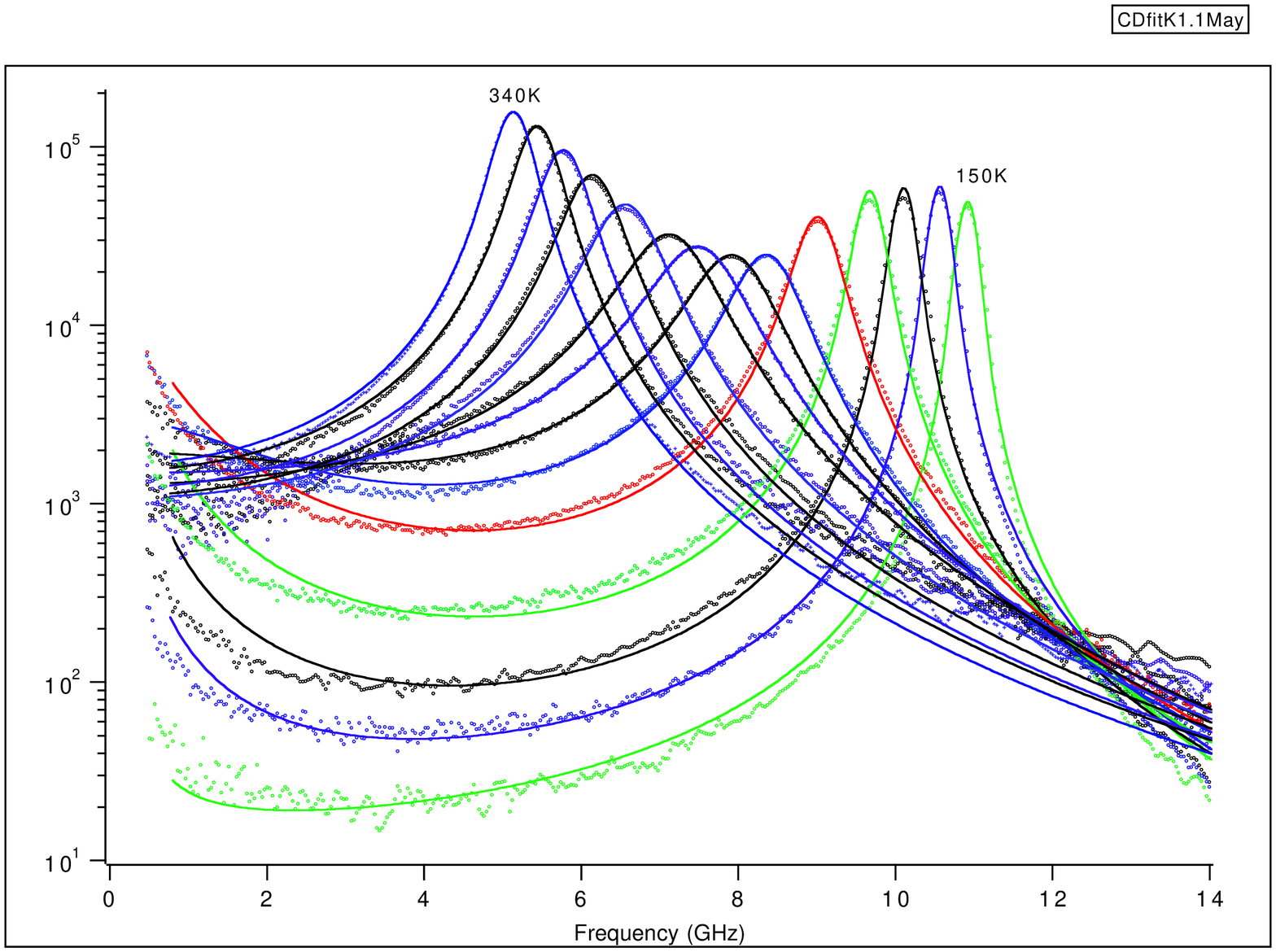}
}
\vspace{0.1in}
}
\refstepcounter{figure}
\parbox[b]{3.3in}{\baselineskip=12pt FIG.~\thefigure.
(CDFit).  Cole-Davidson fits to the $I_{ISO}(\omega)$
spectra of Fig.~3 to Eqs.~(3) and (4), with $\beta = 0.68$,
$\omega_{0}/2\pi = 13.211 - 3. 291 \times 10^{-2}$ T $+ 2.6973 
\times 10^{-5}$ T$^{2}$, $\gamma_{0}/2\pi = T/1000$.
The fits are shown for $T = 150$, 160, 170, 180, 195, 210, 220, 230,
240, 260, 280, 300, 320, and 340~K.}

\vspace{0.10in}
\label{1}
\vspace{0.1in}
  However, the major point of the C.D.\ fits, as shown
by the solid circles in Fig.~1, is that for temperatures below 
$\sim 230$~K, $\tau_{\alpha}^{CD}$ determined from these fits
does not increase very rapidly with decreasing $T$.  At 
$T=T_{G} = 160$~K, $\tau_{\alpha}^{CD}$ found from the C.D.\ 
fits is only $\sim 30$~ns, compared to $\tau \sim 100$~s found in other
experiments.  This disagreement is generally recognized as
indicating that the C.D.\ model (or any $\alpha$-relaxation-only
model) for $m(\omega)$ is incomplete.  Once the $\alpha$ peak
moves well below the frequency of the Brillouin peak,
$m''(\omega)$ of the C.D.\ function [with reasonable $\tau(T)$ values]
decreases rapidly with decreasing temperature, while the shape of the
spectrum indicates that a significant contribution from $m''(\omega)$
is still present in the region of the Brillouin peak.

The fact that the C.D.~function is an incomplete representation of the
structural relaxation dynamics can also be seen directly in the PC dielectric
data of Schneider {\it et al} [Schneider~1999].  Their Fig.~2 shows 
C.D.\ fits to the dielectric data which are excellent for the $\alpha$ peak,
but which fall below the experimental data in the high-frequency wing.

\section{Data Analysis 2: Including the fast $\beta$ relaxation}

The $\tau(T)$ values obtained from the Cole-Davidson fits 
exhibit temperature dependence at low temperatures that 
disagrees strongly with the results of other experiments,
as seen in Fig.~1.  As mentioned in
the introduction, this disagreement is a 
characteristic failure of $\alpha$-relaxation-only
models for $m(\omega)$, which indicates the
need for extending $m(\omega)$ to include
fast relaxation processes.  Several authors have
proposed models that extend the C.D.\ model
empirically by adding another phenomenological
term to Eq.~(4), e.g.~a Cole-Cole term [Soltwisch~1998], 
a damped oscillator [Lebon~1997],
or a Lorentzian [Monaco~2000B]; however,
attributing a fast relaxation time
$\tau_{\beta}$ to structural relaxation is 
incompatible with MCT which
describes the fast relaxation by power 
laws in $\omega$ or $t$ for which
there is no $\tau_{\beta}$ [G\"{o}tze~1992]. 
A major advantage of this
approach is that the additional fast C.D.\ or 
Lorentzian, with relaxation time
$\tau_{\beta} \sim 30$~ps, is able to 
explain the extra ``Mountain mode'' 
seen in OTP at low temperatures 
[Monaco~1999, Fig.~1(a)].  (Recently, 
however, this feature has been attributed
to Mountain's original mechanism
of intramolecular dynamics rather than 
to structural relaxation~[Monaco~2000, Mossa~2001].) 

\subsection{The hybrid model}

We have utilized a different additive memory function,
the hybrid model (mentioned in the introduction), 
which has been employed in several previous Brillouin scattering studies.
We begin with the C.D.\ function [Eq.~(4)] and add a term to
$m(\omega)$ proportional to $\omega^{a-1}$ which represents the
$t^{-a}$ ``critical decay'' part of the fast $\beta$ process in MCT.  This 
superposition approximates the two-step relaxation scenario of MCT,
and does not introduce another relaxation time $\tau_{\beta}$.

In some previous studies, we have used a related procedure to analyze
depolarized backscattering spectra.  For orthoterphenyl, we combined
extended MCT fits for the $\beta$-relaxation region with KWW fits
for the $\alpha$ peak [Cummins~98].  A comparison of these two
hybrid procedures has not yet been attempted.

Two problems arise when the term $t^{-a}$ is added to $m(t)$.  
First, the resulting $\omega^{a}$ term in $\omega m(\omega)$ 
extends to arbitrarily high frequencies.  This is presumably
unimportant for Brillouin scattering, however, since fits to the 
Brillouin spectra are limited to $\omega/2\pi \lesssim 15$~GHz.  Second, 
$m''(\omega)$, which appears in the numerator of $I (\omega)$ in 
Eq.~(2), diverges at low frequencies as $\omega^{a-1}$ resulting 
in the introduction of a spurious ``central peak''.  To eliminate this 
artifact, we have included an exponential cutoff in the critical decay 
part of $m(t)$ [Gotze~2000B]:

\begin{equation}
m_{crit}(t) = B t^{-a}e^{-(t/\tau)} \ 
\end{equation}
where $\tau$ is the same $\tau_{\alpha}^{CD}$ of the C.D.\ function, 
Eq.~(4). (Note that this cutoff factor has not been included in 
previous analyses based on the hybrid model.)

With this modification, the hybrid memory function becomes
\begin{eqnarray}
\omega m(\omega) = \Delta^{2} [(1-i\omega \tau)^{-\beta}-1] \nonumber\\ 
+i \omega B \Gamma(1-a) (\tau^{-1}-i \omega )^{a-1} \ 
\end{eqnarray}
[see the Appendix for the derivation  of Eq.~(7).]

At high frequencies where $\omega \tau >> 1$, the critical decay part
of Eq.~(7) becomes

\begin{equation}
\omega m_{CR} (\omega) = [B_{1} + i B_{2}] \omega^{a} \ 
\end{equation}
\{where $B_{1}=B \sin [ \pi(a-1)/2] \Gamma(1-a)$ and
$B_{2} = B \cos [ \pi (a-1)/2] \Gamma (1-a)$ \}.  Equation~(8) 
is the familiar high-frequency power-law behavior predicted by MCT.

In Fig.~6~ we plot $\omega m''(\omega)$ of Eq.~(7) with
$\tau~=~0.3$~ns, $\beta~=~0.68$, and $a~=~0.29$.  The broken line is 
the Cole-Davidson term only, and the lower solid curve is the 
critical decay term only, including the exponential cutoff.  The upper 
solid curve is the full  hybrid memory function $\omega m''(\omega)$
of Eq.~(7) with $B /\Delta^{2} = 0.005$.  This figure illustrates how
the resulting $\omega m''(\omega)$ goes over smoothly from the C.D.\
$\alpha$-peak behavior at low frequencies to the $\omega^{a}$ behavior
at high frequencies, with a minimum between the two regions.  
The inset to Fig.~6 shows the full 
$m''(\omega)$ with and without  the $e^{-t/\tau}$ cutoff, illustrating  
how the spurious central peak is eliminated
by the exponential cutoff. 

\subsection{The critical exponent $a$}

In MCT, the function $\omega m ''(\omega)$ is predicted to have minimum
between the high-frequency von-Schweidler wing of the
$\alpha$ peak and the critical decay region similar to that in Fig.~6.
In the region of this minimum, the MCT interpolation equation gives 
$\omega m ''(\omega)$ in terms of
the critical exponents $a$ and $b$ by
\begin{eqnarray}
\omega m''(\omega) = \omega m_{min}''(\omega) 
[b(\omega/\omega_{min})^{a} \nonumber\\ 
+a(\omega_{min}/\omega)^{b}]/(a+b) \ 
\end{eqnarray}
where $a$ is the critical exponent and $b$ is the von-Schweidler exponent,
representing the high-frequency wing of the $\alpha$ peak.

Since in our hybrid memory function, Eq.~(7), the C.D. function
for $\omega \tau >> 1$ is $\propto \omega^{-\beta}$, the interpolation
Eq.~(9)  with $b = \beta$ will determine an ``effective'' $a_{eff}$
to use in Eq.~(7) at temperatures high enough for 
the minimum to be observable (from previous MCT fits,
$b \approx 0.50$ while $\beta \sim 0.68$).  At lower
$T$, where the minimum has disappeared from the frequency region
of the Brillouin spectrum, $\omega m''(\omega)$ may have
the form of a power law in $\omega$, but since this region no
longer corresponds to the asymptotic region of the minimum,
the apparent power law behavior of $\omega m''(\omega) \propto
\omega^{a_{eff}}$ can produce an $a_{eff}$ quite different
from the critical exponent $a$ itself.

%
%
\vbox{
\vspace{0.2 in}
\hbox{
\hspace{0in}
\epsfxsize 3.2in \epsfbox{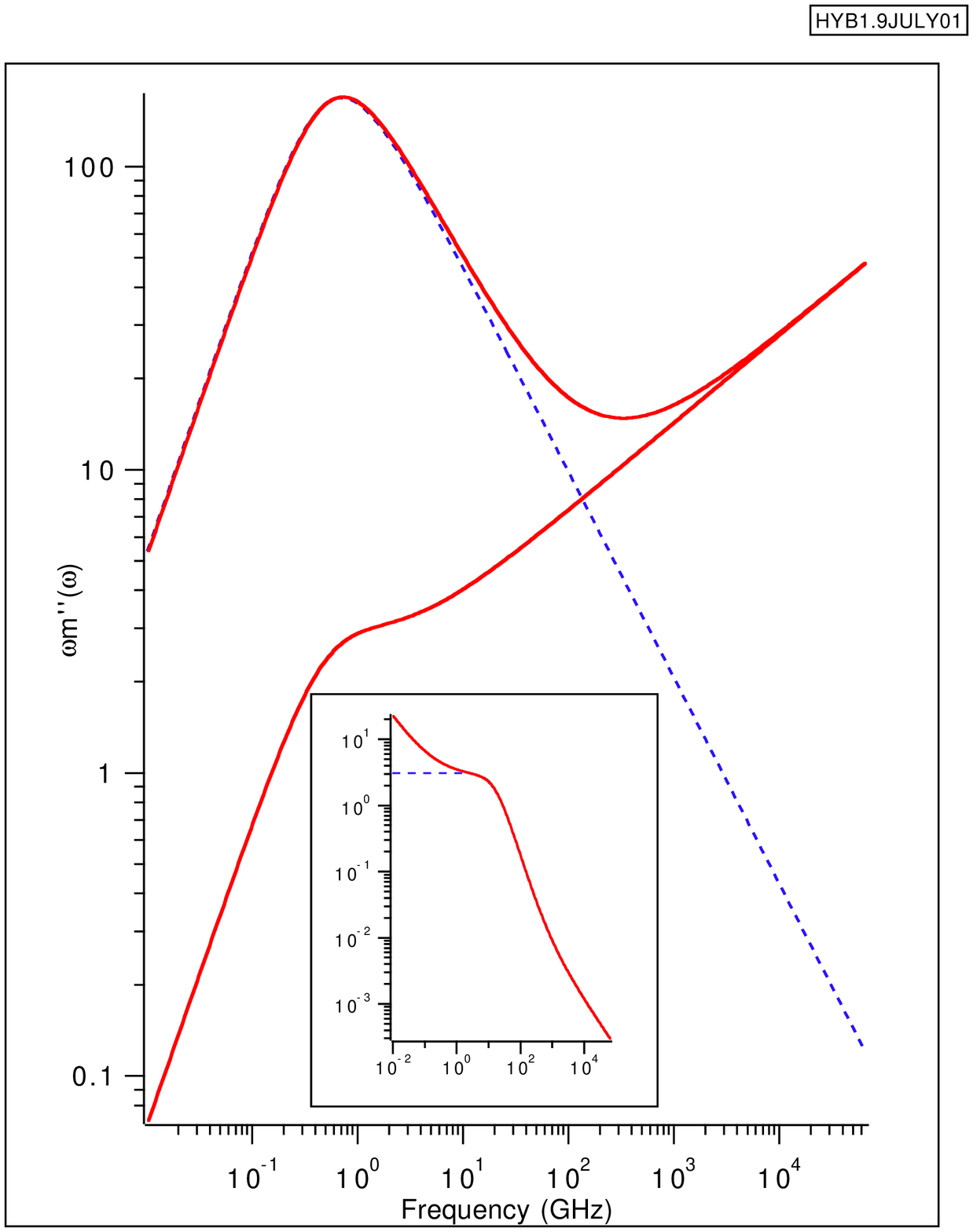}
}
\vspace{0.1in}
}
\refstepcounter{figure}
\parbox[b]{3.3in}{\baselineskip=12pt FIG.~\thefigure.
(HYB1).  The hybrid memory function $\omega m''(\omega)$
of Eq.~(7) with $\tau = 0.3$~ns, $\beta = 0.68$, and $a = 0.29$.
The broken line is the C.D.\ function and the lower solid line is the
critical decay term.  The ratio of the critical decay to C.D.\ 
contributions is $B/\Delta^{2} = 0.005$.  Inset:  $m''(\omega)$ vs
 $\omega$ for the hybrid model [Eq.~(7)] without a cutoff
(solid line), with $\tau = 0.01$~ns, $\Delta = 20$~GHz, and $B = 5$;
$m''(\omega)$ with the cutoff $e^{-t/\tau}$ in Eq.~(6) (broken line).}

\vspace{0.10in}
\label{1}
\vspace{0.1in}

We utilized the depolarized backscattering spectra
$I_{VH}(\omega)$, multiplied by $\omega$ to estimate
$\omega m''(\omega)$ in order to find $a_{eff}(T)$.
In Fig.~7~ we show six $\omega I_{VH}(\omega)$
spectra for $T=140$, 150, 160, 170, 180, and 185~K.
The spectra for $T=170$, 180, and 185~K, where the
minimum is visible, have been fit to the interpolation 
equation [Eq.~(9)] using $\beta = 0.68$ for $b$, while
the spectra for $T=140$, 150, and 160~K have
been fit to $\omega^{a_{eff}}$.  The values of
$a_{eff}(T)$ found from these fits are:

\begin{center}
\begin{tabular*}{1.25in}{cc}
\makebox[1in]{\bf $T$} &  \makebox[1in]{\bf $a_{eff}$} \\
\makebox[1in]{140}  &  \makebox[1in]{0.62}  \\
\makebox[1in]{150}  &  \makebox[1in]{0.59}  \\
\makebox[1in]{160}  &  \makebox[1in]{0.52}  \\
\makebox[1in]{170}  &  \makebox[1in]{0.42}  \\
\makebox[1in]{180}  &  \makebox[1in]{0.30} 
\end{tabular*} 
\end{center}
In the fits to our Brillouin spectra, we used these values of
$a_{eff}(T)$ for temperatures up to 180~K and $a = 0.29$ for all higher
temperatures.
%
%
\vbox{
\vspace{0.2 in}
\hbox{
\hspace{0in}
\epsfxsize 3.2in \epsfbox{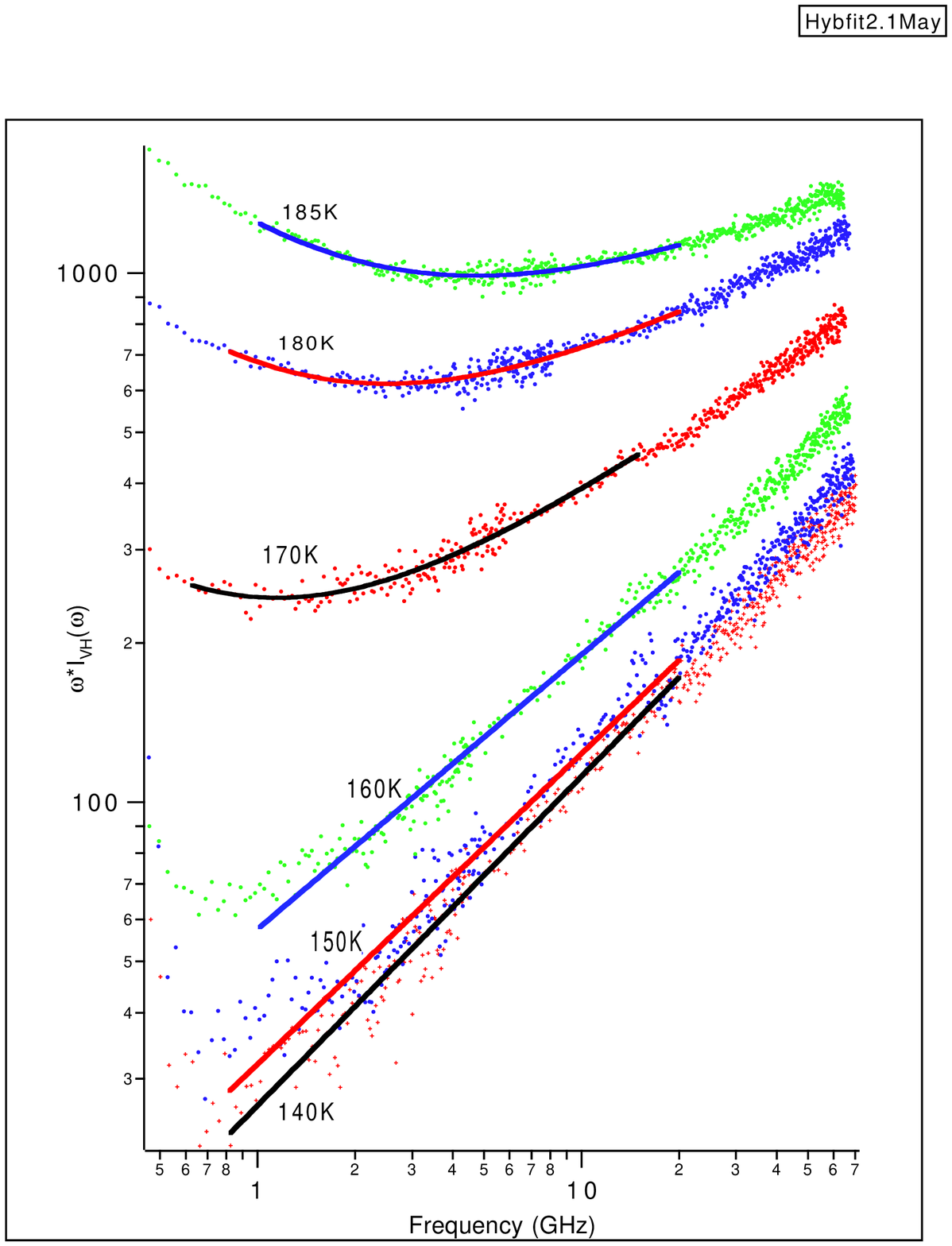}
}
\vspace{0.1in}
}
\refstepcounter{figure}
\parbox[b]{3.3in}{\baselineskip=12pt FIG.~\thefigure.
(HYBFIT2).  VH backscattering spectra multiplied by
frequency with fits to the MCT interpolation equation
[Eq.~(9)] (using $b=\beta = 0.68$) ($T = 170$, 180, 185~K)
or the power law $\omega^{a_{eff}}$ $(T = 140$, 150, 160~K).
The values of $a_{eff}$ determined in these fits were used
in the low-temperature $I_{ISO}(\omega)$ hybrid model fits 
shown in Figs 8 and 9. 

\vspace{0.10in}
}
\label{1}
\vspace{0.1in}
\subsection{Minimum free parameter analysis}

We initially tried to fit the spectra with Eq.~(3) and the hybrid memory
function Eq.~(7) with the smallest possible number of free parameters. 
We fixed  $\omega_{0}(T)$ with Eq.~(5), fixed  $\gamma_{0}(T) =
T/1000$ as in the Cole-Davidson fits, fixed $\tau$ using Eq.~(1), fixed
$\beta~=~0.68$, and fixed the ratio $B/\Delta^{2}$, the 
relative strengths of the critical decay term ($B$) and Cole-Davidson
term ($\Delta^{2}$), by the following strategy. For fixed $\tau$, the
location of the minimum in the function $\omega m''(\omega)$ 
occurs at a frequency $\omega_{min}$ that decreases as 
$B/\Delta^{2}$ increases. We then invoked the MCT result that 
$\omega_{min}$ (unlike $\tau_{\alpha}$) should be equivalent for 
all variables, and we adjusted $B/\Delta^{2}$ to make $\omega_{min}$
fall close to the $\omega_{min}$ found in the depolarized backscattering
PC $\chi ''(\omega)$ spectra of Du {\it et al} [Du~1994].  
For $T=190$, 210, 230, 250 and 270~K, the $\omega_{min}$ values of 
Du~{\it et al} were found to agree quite well with the minima of the 
$\omega m''(\omega)$ curves computed with Eq.~(7) with 
$B/\Delta^{2} = 0.175$.  With this fixed value of $B/\Delta^{2} = 0.175$,
there are only two free parameters left in Eqs.~(7) and (3): 
the C.D.\ coupling constant $\Delta^{2}$, and the overall normalization
constant ${I_{0}}$.

With this severely constrained fit procedure, however, the resulting fits 
were not generally acceptable. One surprising aspect of the results was 
that for $T \leq 170$~K the linewidths of the fits became too broad; this
problem can be resolved by {\em decreasing} $B/\Delta^{2}$. 

This indication that the strength of the critical contribution to the
memory function (relative to the $\alpha$ peak) may weaken in
the temperature range between $T_{C}$ and $T_{G}$ is also
supported by the depolarized backscattering spectra, shown in
Fig.~7.  At temperatures below $T = 170$~K, the minimum
in $\omega I (\omega)$ is at $\omega \leq 1$~GHz, so the
$\alpha$ peak is completely out of the spectral window.  The intensity in
the Brillouin frequency region ($\sim 8$~GHz), which is therefore entirely
due to the ``fast beta decay'', is seen to decrease rapidly with
decreasing $T$, consistent with the apparent behavior of
$\omega m (\omega)$.  This result is also presumably related to the
MCT prediction of a cusp in the non-ergodicity parameter $f_{q}(T)$
at $T_{C}$ below which the critical decay weakens with decreasing
temperature.  It also indicates that additional contributions to the fast
relaxation from other processes (e.g.~intramolecular vibrations) are
unlikely in PC since these would produce a larger value of $B/\Delta^{2}$.

\subsection{Complete hybrid model fits}

To provide a complete hybrid model analysis, we allowed the ratio
$B/\Delta^{2}$ to vary in the fit. At low temperatures, $B$ and
$\Delta^{2}$ could be determined independently from the fit. At 
temperatures above 200~K, $B$ and $\Delta^{2}$ are strongly 
correlated. At 210~K an optimum fit was found with $B/\Delta^{2} = 4.0$,
and we arbitrarily kept this value fixed for all temperatures above 210~K.
The resulting set of fits is shown in Fig.~8; the fitting parameters
$B/\Delta^{2}$ and $\tau$ are given in Table~II together with the 
$\chi^{2}$ values. These hybrid model fits have three free parameters,
as in the C.D.\ fits.   However, they have $\tau(T)$ constrained to follow 
Eq.~(1) shown by the broken line in Fig.~1, automatically 
eliminating the $\tau(T)$ disagreement problem of  the C.D.\ fits.  
They are nevertheless at least as  good as the C.D.\ fits. 

In their light scattering study of meta-toluidine, Aouadi {\it et al} 
[Pick~2000] also analyzed $I_{ISO}(\omega)$ spectra with the
hybrid model, constraining $\tau_{\alpha}^{CD}$ to be proportional
to the $\tau_{\alpha}$ determined from depolarized backscattering
spectra.  They found, as we did, that at low temperatures the strength of the
critical decay decreases with decreasing $T$ (see their Fig.~12).  Although
they also found that the critical decay strength decreases at high temperatures,
resulting in an apparent maximum around 250~K, this behavior may be a
manifestation of the strong correlation between $\Delta^{2}$ and $B$
mentioned above.
%
\vbox{
\vspace{0.2 in}
\hbox{
\hspace{0in}
\epsfxsize 3.2in \epsfbox{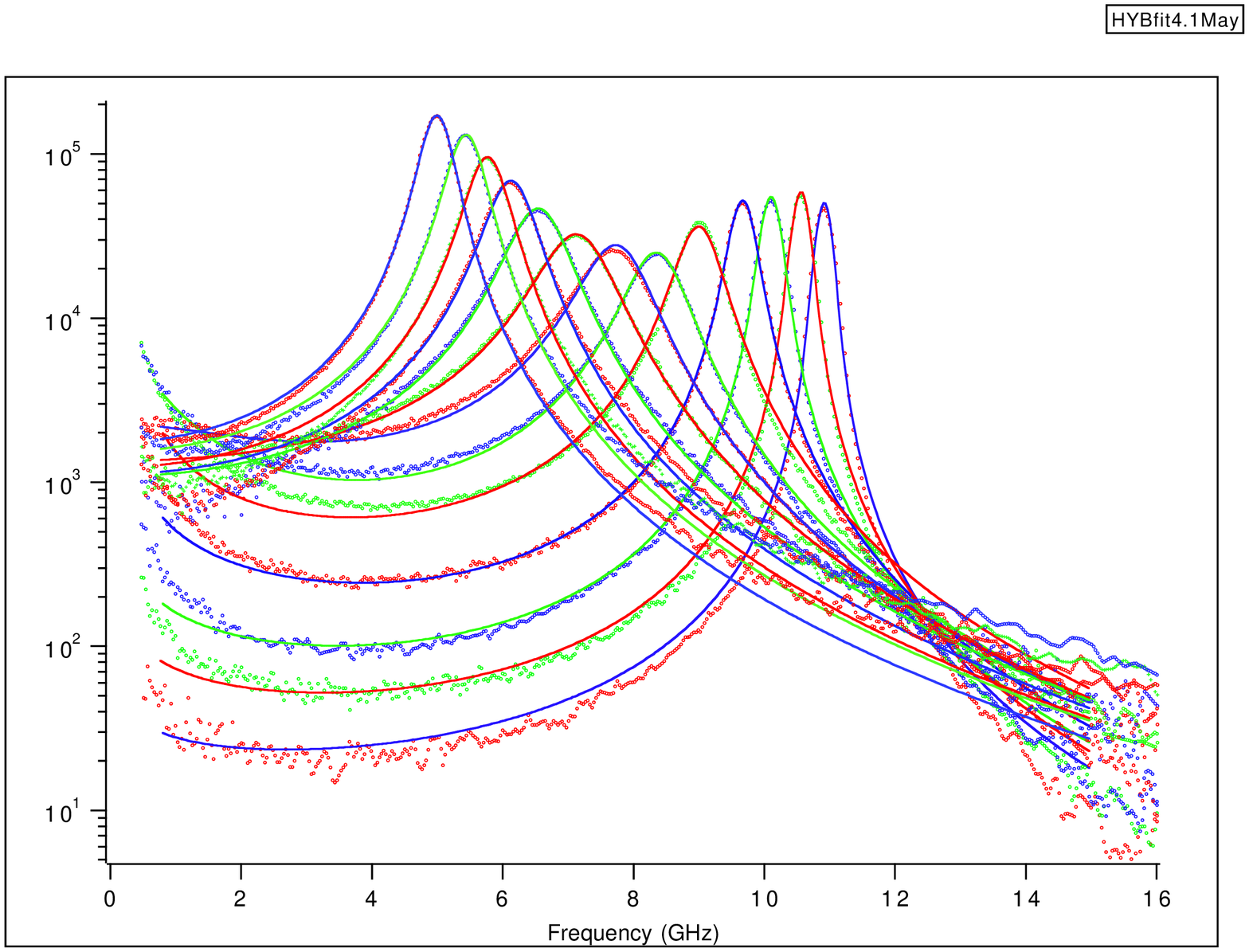}
}
\vspace{0.1in}
}
\refstepcounter{figure}
\parbox[b]{3.3in}{\baselineskip=12pt FIG.~\thefigure.
(HYBFIT4). 
  Fits to PC Brillouin spectra $I_{ISO}(\omega)$
with the hybrid model for $m(\omega)$ of Eq.~(7).  The free fitting
parameters are $I_{0}, \Delta^{2}$, and $B/\Delta^{2}$; the values
found from the fits are given in Table~ II.  For $T > 210$ K, $\Delta^{2}$
and $B/\Delta^{2}$ were too strongly correlated to be determined by the fit, 
so $B/\Delta^{2}$ was arbitrarily fixed at 4.0.}
\vspace{0.10in}

\label{1}
\vspace{0.1in}

\subsection{Thermal diffusion contribution}

With  $a_{eff}$ replacing $a$ in Eq.~(7), we carried out additional
fits to the Brillouin spectra for $T=150$, 160, 170, and 180~K.  
The fits are shown by the lower curves for each temperature in 
Fig.~9. (They are also the four low-$T$ fits shown in Fig. ~8~).  
Note that at low temperatures the experimental
spectra have weak extra structure at low frequencies which is not 
present in the theoretical fits. 

The light scattering spectrum of density fluctuations
$I_{ISO}(\omega)$, including thermal diffusion effects,
can be expressed as [Boon~1980], [Scopigno~2000]
\begin{equation}
I(\omega) = \frac{I_{0}}{\omega} {\cal I}m \left[
\omega_{T}^{2} - \omega^{2} - i \omega \gamma_{0}
- \omega m(\omega) \right]^{-1} \ \ 
\end{equation}
where $\omega_{T} = C_{T}q$ ($C_{T}$ is the
isothermal sound velocity) and 
$m(\omega)= m_{\eta}(\omega) + m_{Th}(\omega)$
includes both the structural relaxation contribution
$m_{\eta}(\omega)$ [e.g.~Eqs.~(4) or (7)] and
the thermal diffusion contribution
\[ m_{Th}(t) = 
(\gamma - 1)\omega_{T}^{2} e^{-t/\tau_{Th}} \]
where $\tau_{Th} = (D_{T}q^{2})^{-1}$.
With the Laplace transform convention of Eq.~(18) in the Appendix,
\begin{equation}
m_{Th}(\omega) = \frac{i (\gamma - 1) \omega_{0}^{2} 
\tau_{Th}}{1- i \omega \tau_{Th}} \ \ 
\end{equation}
In the spectral window explored by $90^{\circ}$
Brillouin scattering ($\omega \gtrsim 0.5$~GHz),
$\omega \tau_{Th} >> 1$ since typically $\tau_{Th}
\sim 15$~ns.  Then, in Eq.~(10), $-\omega m_{Th}(\omega)
\sim \omega_{T}^{2} (\gamma - 1)$.  Adding this to the
$\omega_{T}^{2}$ term in Eq.~(10) then gives
$\omega_{T}^{2} + \omega_{T}^{2} (\gamma - 1) =
\omega_{T}^{2} \gamma$ which is just the square of the
adiabatic sound frequency $\omega_{0}$.  In that
approximation Eq.~(10) recovers the simpler
equation [Eq.~(3)] with $m( \omega) \equiv
m_{\eta}(\omega)$ only.

While the quasielastic thermal diffusion mode is not
usually included in the analysis of Brillouin scattering
spectra, it {\em is} included in the analysis of inelastic x-ray scattering
experiments since with larger $q$ it becomes much broader
[Scopigno~2000].  Also, at low temperatures, the
high-frequency tail of the thermal diffusion mode
can appear in the Brillouin spectrum if the window
extends to sufficiently low frequencies.  The four
spectra shown in Fig.~9 have weak
extra structure at low frequencies which may be due to 
the tail of the thermal diffusion mode.

Since numerical data for $\gamma$ and $D_{T}$ are not
available for PC (to our knowledge), we estimated
$\gamma = 1.5$ and $\tau_{Th} = 16$~ns, and
reanalyzed the spectra of Fig.~9~ using
Eq.~(10) with $\omega_{T}^{2} = \gamma \omega_{0}^{2}$,
$m(\omega) = m_{\eta}(\omega) + m_{Th} (\omega)$
with $m_{Th}(\omega)$ given by Eq.~(11), and
$m_{\eta}(\omega)$ by Eq.~(7).  The resulting fits,
including thermal diffusion, are shown as the upper curves
for each temperature in Fig.~9~ for $T = 150$, 160, and 170~K,
showing that at these low temperatures, the thermal diffusion mode
is apparently visible in the Brillouin spectrum at low frequencies.  
For temperatures above 170~K, the thermal diffusion contribution is not visible.

\subsection{Hybrid model with $\beta = b$}

Finally, we note that a slightly different strategy can be followed
in carrying out the hybrid model analysis.  With the exponent $a$ in
Eqs.~(6) and (7) fixed (for $T > T_{C}$) at 0.29, the region 
of the minimum in $\omega m''(\omega)$ is predicted by MCT to
obey Eq.~(IV.5) with $b = 0.50$.  However, there is a slight 
difference between this prediction and Eq.~(7) because so far 
we have taken $\beta \neq b$.  This difference can be eliminated by 
arbitrarily fixing $\beta = b = 0.50$, as was done for the analysis of 
depolarized backscattering spectra of toulene by
Wiedersich~{\it et al}~[Wiedersich~2000].
  
We therefore carried out another set of hybrid model fits with
$\beta=b=0.5$.  All other parameters were the same as in the fits
of Fig.~8.  In these fits, we kept $a=0.29$ and $\beta = 0.50$
for all temperatures and also kept both $\Delta^{2}$ and $B$ free.
The resulting fits are comparable to those
obtained with $\beta = 0.68$, demonstrating that the hybrid model
fits are not very sensitive to the value of $\beta$.
%
\vbox{
\vspace{0.2 in}
\hbox{
\hspace{0in}
\epsfxsize 3.2in \epsfbox{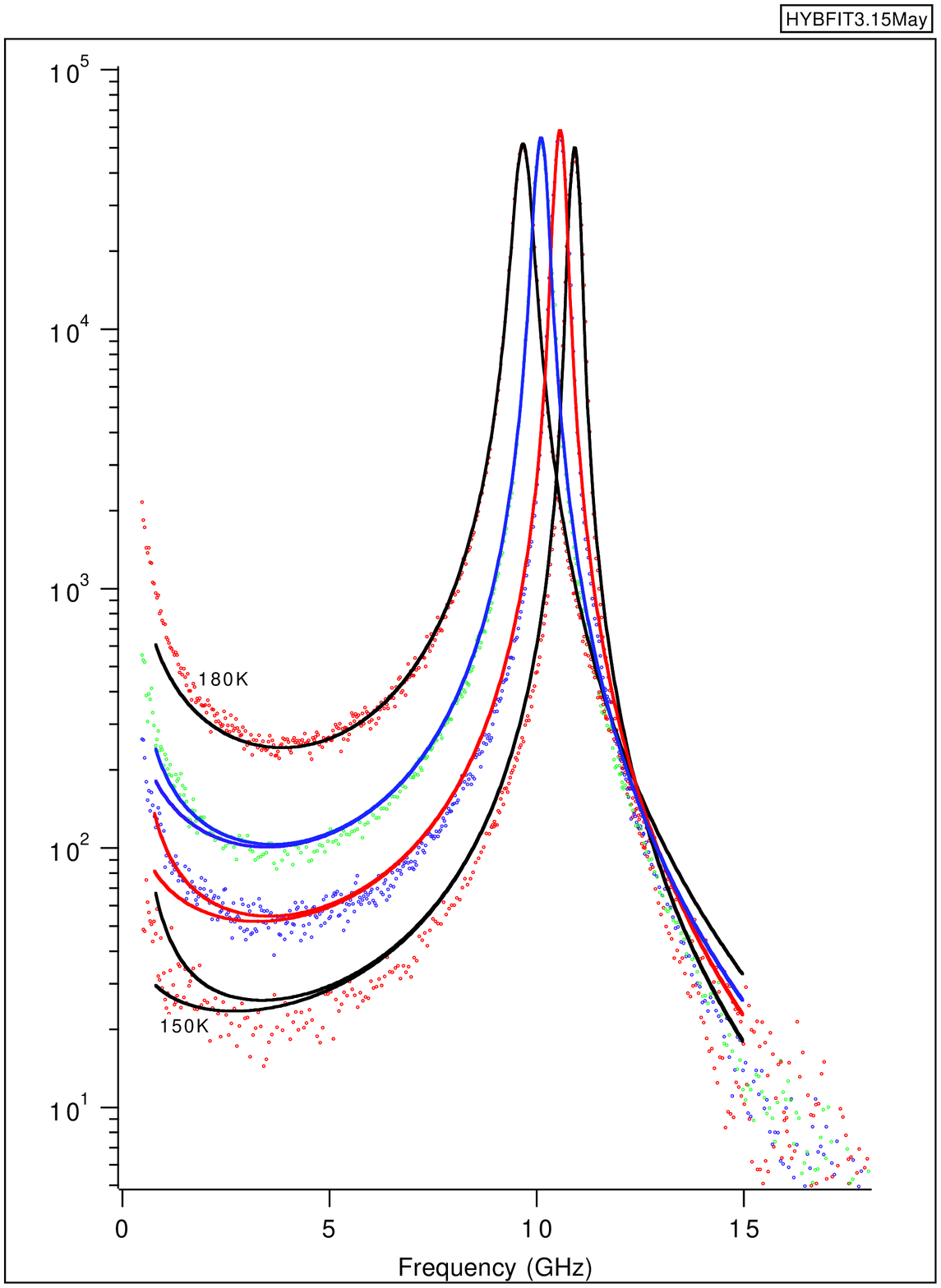}
}
\vspace{0.1in}
}
\refstepcounter{figure}
\parbox[b]{3.3in}{\baselineskip=12pt FIG.~\thefigure.
(HYBFIT3).  Low-temperature hybrid model fits using the
$a_{eff}$ values found in the fits shown in Fig.~7.
For $T = 150$, 160, and 170~K, a second fit (the upper curve
at each temperature) is included for each spectrum which includes
the effect of thermal diffusion.}
\vspace{0.10in}

\label{1}
\vspace{0.1in}

  At low temperatures,
however, the low-frequency regions of these fits are not as good as
the previous fits, illustrating the effect of the increase in $a_{eff}$  at
low $T$.  The $\chi^{2}$
values obtained both with $\beta = 0.68$ and $\beta = 0.5$ are 
listed in Table~II together with the $B/\Delta^{2}$ values, and
are seen to be similar.  Note that in the $\beta=0.5$ fits where
$\Delta^{2}$ and $B$ were both free fitting parameters,
$B/\Delta^{2}$ increases with increasing $T$, passes through a
maximum at $T \sim 230$~K, and then decreases again at high $T$,
similar to the results of Aouadi~{\it et al} for meta-toluidine~[Pick~2000].
As noted above, however, $B$ and $\Delta^{2}$ are strongly correlated
at high temperatures, and the 
high-$T$ values of $B/\Delta^{2}$ are therefore not significant.

An advantage of choosing $\beta=b$ is that there is one less free
parameter in the fitting procedure.  Also, the region of the susceptibility
minimum in Eq.~(9) is the same as in Eq.~(7) if $\beta = b$.
This makes the $\beta = b$ hybrid model a good candidate for attempting
to simultaneously describe all relaxing properties in a given material.

\section{Data Analysis 3: Extended Schematic MCT model}

For the third and final analysis of the Brillouin spectra 
$I_{ISO}(\omega)$, we constructed the memory function
$m(\omega)$ of Eq.~(3) from the mode coupling theory 
rather than from parameterized empirical models, using the extended
schematic MCT model of G\"{o}tze and Voigtmann mentioned in the
Introduction [G\"{o}tze 2000].

\subsection{Mode coupling theory}

Mode coupling theory (MCT) begins with exact classical equations of
motion (the generalized Langevin equation) for $\phi_{q}(t)$, 
the normalized autocorrelation functions of the density fluctuations 
$\rho_{q}(t)$, derived from the Hamiltonian with the Zwanzig-Mori
projection-operator formalism [Gotze~1992]:

\begin{equation}
\ddot{\phi}_{q}(t) + \nu\dot{\phi}_{q}(t) + \Omega_{q}^{2}\phi_{q}(t) +
\Omega_{q}^{2} \int^{t}_{0} m_{q}(t-t')\dot{\phi}_{q}(t') dt'=0 \ \ 
\end{equation}

Equation (12) is equivalent to the generalized hydrodynamics equation
(3), as shown in the Appendix, but its parameters are determined by the Hamiltonian in the
Zwanzig-Mori formalism.  In particular, $m_{q}(t-t')$, which is an
empirical memory function in Eq.~(3), is the autocorrelation
function of the random forces in Eq.~(12).  A central accomplishment
of MCT was the derivation of a closed mode-coupling approximation
for $m_{q}(t)$ as

\begin{equation}
m_{q}(t) = \sum_{q_{1}} V(q,q_{1},q-q_{1})\phi_{q_{1}}(t) \phi_{q-q_{1}} (t) \ 
\end{equation}
in which the coupling coefficients $V(q,q_{1},q-q_{1})$ are expressed in
terms of the static structure factors $S_{q}$, $S_{q_{1}}$, and $S_{q-q_{1}}$.
If the intermolecular potential is known, Eqs.~(12) and (13) can be
solved self-consistently to find the density correlation functions and
spectra [Franosch~1997b].

In cases where the intermolecular potential is not known, simplified
``schematic'' MCT models can be used which include only a few correlators,
with the remaining coupling constants considered as fitting parameters.
The Sj\"{o}gren model is a two-correlator schematic model.  The ``system'' is
represented by the system correlator $\phi(t)$, while the variable
being probed experimentally is represented by the probe correlator
$\phi_{s}(t)$.  The two memory functions are
\begin{mathletters}
\begin{equation}
m(t) = v_{1}\phi(t) + v_{2}\phi^{2}(t) \ \ \ 
\end{equation}
\begin{equation}
m_{s}(t) = V_{s}\phi(t)\phi_{s}(t) \ \ \ 
\end{equation}
\end{mathletters}
Equation (12), written separately for $\phi(t)$ and $\phi_{s}(t)$
with the memory function of Eqs.~(14), can then be solved 
self-consistently with $v_{1}$, $v_{2}$, and
$V_{s}$ as fitting parameters (see [Singh~1998]).

For temperatures near and below $T_{C}$ of MCT, the ``cage effect''
dynamics embodied in Eq.~(12) lead to the (usually) non-physical result
of total structural arrest.  In most glass-forming materials, this arrest is avoided
by activated hopping processes which are included in the extended version
of MCT (to a first approximation) via a temperature-dependent hopping 
parameter $\Delta$.  Equation~(12) is replaced by [G\"{o}tze~2000]
\begin{eqnarray}
\ddot{\phi}(t) + (\Delta +\nu)\dot{\phi}(t) + (\Omega^{2}+\Delta \cdot \nu)
\phi(t) \nonumber\\
+ \Omega^{2} \int^{t}_{0} m(t-t') [\dot{\phi}(t')+ \Delta \cdot \phi
(t')]dt'=0 \ \ 
\end{eqnarray}
(Note that the hopping parameter $\Delta$ in Eq.~(15) is not
related to the coupling constant $\Delta^{2}$ of the Brillouin fits
in, e.g., Eq.~(7).)  

The extended Sj\"{o}gren model of G\"{o}tze 
and Voigtmann consists of two Eqs.~(15), one for $\phi$ and one 
for $\phi_{s}$, with the two memory functions of Eqs.~(14).  
These are to be solved self-consistently.

The parameters of $\phi(t)$ are: $\Omega$, $\nu$, $\Delta$, 
$v_{1}$, and $v_{2}$.  They have already been globally optimized
for PC by G\"{o}tze and Voigtmann for dielectric, neutron scattering
and depolarized light scattering data covering a much larger range of 
frequencies than that of our Brillouin scattering spectra.  The parameters
of $\phi_{s}(t)$ are:  $\Omega_{s}$, $\nu_{s}$, $\Delta_{s}$, and 
$V_{s}$.  (Note that varying the parameters of $\phi_{s}$ has no effect
on $\phi$.)

In this section, the G\"{o}tze-Voigtmann schematic MCT model 
will be used to compute the memory function $m(\omega)$ of Eq.~(3), 
taken as $m(\omega) = \Delta^{2} \phi_{s}(\omega)$, where $\Delta^{2}$
is the coupling constant.  [We note that this is the same procedure employed 
by Ruffl\'{e}~{\it et al} [Ruffl\'{e}~1999] who used the simpler 
Sj\"{o}gren model)]. 

\subsection{MCT fitting procedure}

Fitting the Brillouin spectra $I_{ISO}(\omega)$ with the two-correlator
MCT model described above is technically more difficult and
therefore less straightforward than the two other fitting procedures.
This is because $m(\omega)= \phi_{s}(\omega)$ must first be obtained
by solving and Fourier transforming the MCT equations [Eqs.~(15)], and then
interpolated to match the frequencies of the $I_{ISO}(\omega)$ spectrum.

For the fits with the schematic MCT model, we began with the 
parameters obtained in the depolarized light scattering (DLS) fits of 
G\"{o}tze and Voigtmann, shown in their Figure~1 [G\"{o}tze~2000].  
Since the correlators $\phi_{s}$ computed with the parameters
they used gave very good fits to the depolarized light scattering spectra
over nearly four decades in frequency, we tried to fit the Brillouin 
scattering spectra using these same parameters.  However, the
 $\alpha$ peaks of the DLS spectra are at lower frequencies
than the corresponding $\tau_{\alpha}^{LA}$ values, as seen in 
Fig.~1, where 
$\tau_{\alpha}^{DLS} \sim 5 \times \tau_{\alpha}^{LA}$.  We found
that the position of the $\alpha$ peak in $\omega \phi_{s}''(\omega)$ is 
primarily determined by the coupling constant $V_{s}$ in Eq.~(14b).  We
therefore treated $V_{s}$ as a free fitting parameter.  All other MCT
parameters were kept at their DLS values.  We solved the MCT equations
[Eqs.~(15)] for selected values of $V_{s}$ and the resulting 
$\phi_{s}(\omega)$, after multiplication by the (adjustable) coupling constant
$\Delta^{2}$, was taken as the memory function $m(\omega)$ in Eq.~(3).
A nonlinear least-squares analysis was then carried out in which $I_{0}$ and
$\Delta^{2}$ were the free parameters.  This process was repeated until the fit
was optimized.  (Note that in fitting their Brillouin spectra of NaLiPO$_{3}$
with the idealized two-correlator schematic Sj\"{o}gren model, Ruffl\'{e} {\it et al} 
also treated $V_{s}$ as a free parameter while the system parameters were fixed
from neutron scattering results [Ruffl\'{e} {\it et al} 1999].)
%

The MCT fits for $T=160$, 180, 200, 220, 240, 260, 280, 300, 320,
and 350~K are shown in Fig.~10.  (For $T~\geq~300$~K,
the MCT parameters other than $V_{s}$ were obtained by extrapolation
from lower temperatures.)  The values $V_{s}$ and $\chi^{2}$ for these
fits are given in Table~II (note that the $V_{s}$ values for the Brillouin
fits are systematically smaller than the depolarized light scattering values
$V_{s}^{DLS}$).  The MCT fits are comparable to the
hybrid fits, and have lower $\chi^{2}$ values at some temperatures.
For the lowest temperatures, the low-frequency region of the MCT fits 
is poor, reflecting the known limitations of the model for low frequencies at
$T<T_{C}$ [G\"{o}tze~2000b]. At temperatures above 280~K, the fits were not
sensitive to the value of $V_{s}$ for $V_{s} \lesssim 3.5$.
This is because the $\alpha$ peak in $\omega m''(\omega)$ is then
above the Brillouin line, and the shape of the memory function in the
Brillouin range (1 to 15~GHz) is nearly independent of $V_{s}$.

For the temperatures 180, 200, 220, 240, and 260~K, the position of
the $\alpha$ peak in $\omega m''(\omega)$ was sufficiently
well-determined by the fits to estimate the MCT $\alpha$-relaxation
time $\tau_{\alpha} = 1/\omega_{\alpha}$ from the frequency
of the $\alpha$ peak.  The values found are listed in Table~II and are 
also shown in Fig.~1 by the $\ast$ symbols.  Note that the MCT 
values follow the same qualitative temperature dependence as the dielectric
data, and are similar to (though somewhat lower than) the values we assumed 
for the hybrid fits.  Thus, the MCT approach, in contrast to the other approaches
followed in the preceding two sections, is able to provide reasonable estimates
of $\tau_{\alpha}^{LA}$ purely on the basis of fits to the Brillouin spectra.

These results show that this extended schematic MCT model is able to
simultaneously describe the spectra and also give reasonable fits for
$\tau_{\alpha}(T)$, thus answering the question that originally motivated
this study affirmatively.

%
\vbox{
\vspace{0.2 in}
\hbox{
\hspace{0in}
\epsfxsize 3.4in \epsfbox{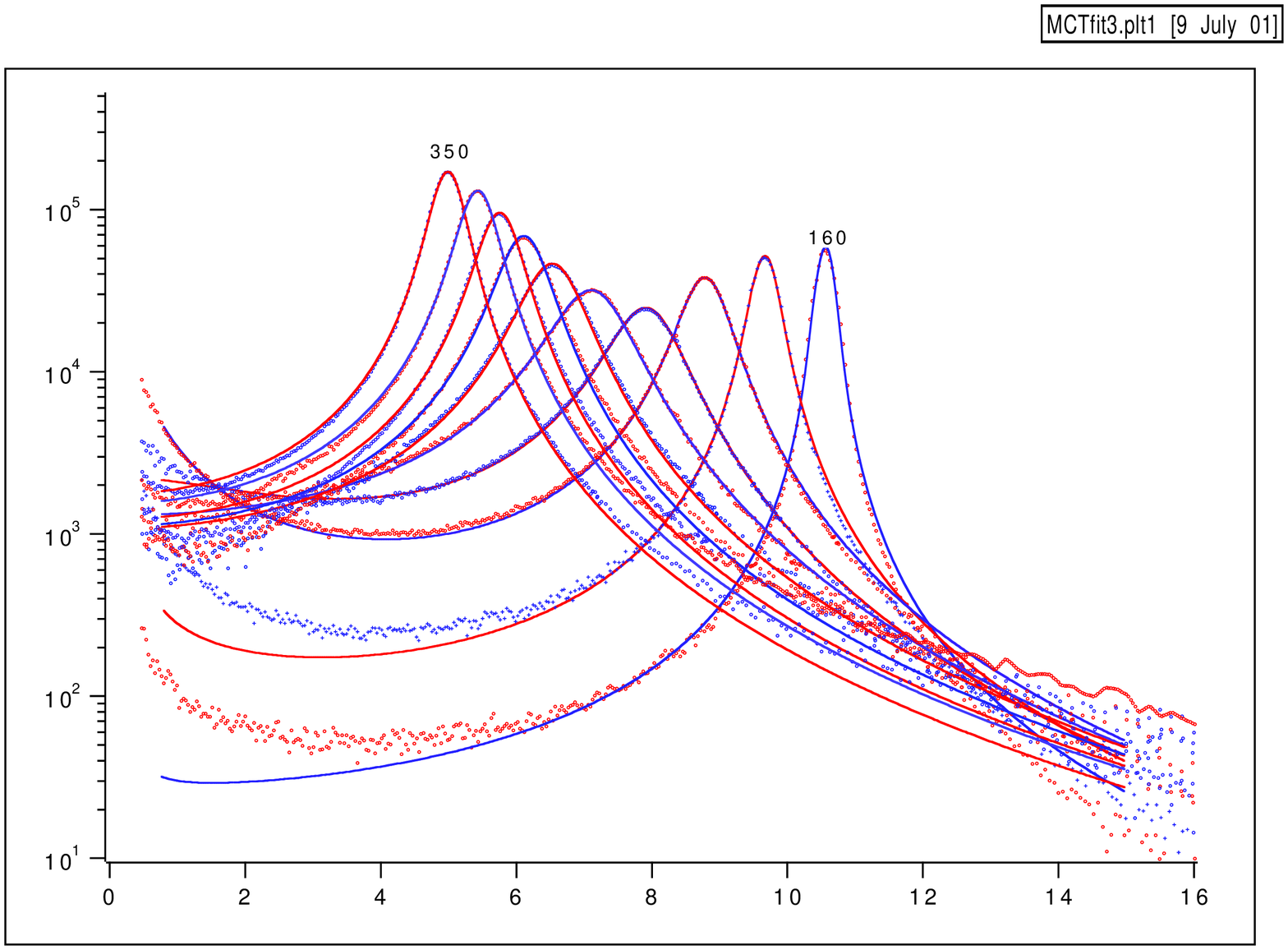}
}
\vspace{0.1in}
}
\refstepcounter{figure}
\parbox[b]{3.3in}{\baselineskip=12pt FIG.~\thefigure.
(MCTFIT3).  Fits of PC $I_{ISO}(\omega)$ spectra 
with the extended schematic MCT model of Eqs.~(15) and (14) for 
$T= 350$, 320, 300, 280, 260, 240, 220, 200, 180, and 160~K.  In the fits, 
$\gamma_{0}(T) = T/1000$ and $\omega_{0}(T)$ is fixed following 
Eq.~(5).  All MCT parameters are those found by G\"{o}tze and 
Voigtmann from their fits to DLS spectra {\em except} for $V_{s}$ 
which has been adjusted to optimize the fits. For $T \geq 300$ K, the
parameters other than $V_{s}$ were determined by extrapolation from
lower temperatures.}
\vspace{0.10in}

\label{1}
\vspace{0.1in}
\section{Discussion and Conclusions}

As noted over 30 years ago by Montrose {\it et al}, the detailed
shape of the $I_{ISO}(\omega)$ Brilliouin spectrum is relatively 
insensitive to the exact nature of the memory function [Montrose~1968].
This is due, as noted earlier, to the limited frequency range of Brillouin
scattering data.  If $\omega_{0}$ is fixed (from ultrasonic data),
the ratio $m''(\omega_{B})/m'(\omega_{B})$ must have the ``correct''
value to obtain reasonable fits, but there is still enormous flexibility in
constructing $m(\omega)$.  If one wishes to extract meaningful information
from the fits, then additional constraints must be imposed.

First, the $\alpha$ relaxation time should be reasonable.  Fits using
$\alpha$-relaxation-only models, e.g.~the Cole-Davidson function of
Eq.~(4), produce acceptable fits but give unreasonably short relaxation 
times at low temperatures.  At $T \sim T_{G}$, one finds typically
$\tau_{\alpha}^{CD}(T_{G})~\sim~10^{-7}$~sec while other 
experimental techniques generally yield $\tau_{\alpha}(T_{G})
\sim 10^{2}$~sec.  This result demonstrates the fact, already shown 
many times, that $\alpha$-only models of structural relaxation are incomplete.

A great deal of research during the past 15 years has shown that 
structural relaxation occurs in two steps, starting with a fast decay
towards a plateau that precedes the final $\alpha$-relaxation process.
Since the memory function $m(\omega)$ for longitudinal viscosity
should also have this two-step form, the memory function used in
the analysis of Brillouin scattering spectra should include both the
$\alpha$-relaxation and the fast-relaxation parts of $m(\omega)$.  If 
$\tau_{\alpha}$ is constrained to follow the temperature dependence 
found in other measurements, such as dielectric or depolarized light
scattering spectroscopy, then the memory function must be extended
to higher frequencies to provide sufficient damping once the $\alpha$
peak in $\omega m''(\omega)$ has moved below the Brillouin lines.
This extension to include fast relaxation processes can be implemented
many ways, but most approaches require the introduction of a fast 
$\beta$ relaxation time $\tau_{\beta}$ that does not correspond to the
results of other experiments.

In practice, there may also be other sources of fast relaxation -- such as
intramolecular vibrations -- that influence the Brillouin spectra, but the
high-frequency part of the structural relaxation dynamics must still be
present.

Because the mode coupling theory has been successful in analyzing a wide
range of experimental data, we have used an empirical model for $m(\omega)$
which is qualitatively consistent with MCT.  The hybrid model, which mimics
the two-step relaxation scenario of MCT by combining the C.D.\ function for
$\alpha$ relaxation with the critical decay law ($\omega^{a}$) has been shown
to provide generally excellent fits to the Brillouin scattering spectra.

Furthermore, we found that fixing $\beta = b = 0.5$ produces fits of 
equally good quality as with $\beta = 0.68$.  This reduction in the number 
of free fitting parameters represents an attractive variant of the hybrid model
which has been shown, for the case of toluene, to provide good fits to
the depolarized backscattering spectra [Wiedersich 2000].  It will be
interesting to explore the ability of this reduced hybrid model to
simultaneously fit both depolarized backscattering spectra and Brillouin
spectra for any given material.

We have also explored the applicability of the two-correlator schematic
MCT model to the analysis of Brillouin spectra, a procedure first employed
by Ruffl\'{e} {\it et al} [Ruffl\'{e}~1999] who used the Sj\"{o}gren model
without hopping.  We used the new extended model of G\"{o}tze and 
Voigtmann~[G\"{o}tze~2000] with which they fit depolarized light scattering, 
neutron scattering, and dielectric data for propylene carbonate
with common parameters for the system correlator $\phi(t)$ for all
the experiments.  For the
probe correlator $\phi_{s}(t)$, we kept the parameters $\Omega_{s}$,
$\nu_{s}$, and $\Delta_{s}$ they obtained from fits to depolarized
light-scattering spectra fixed, but we varied $V_{s}$ to optimize the fits
to the Brillouin spectra, taking $m(\omega) = \Delta^{2} \phi_{s}(\omega)$
where $\Delta^{2}$ is a free parameter.  The resulting $V_{s}$ values
were systematically smaller [and the resulting frequency of the $\alpha$
peak in $\omega m''(\omega)$ higher] than those found from fits to the
depolarized light scattering spectra, consistent with the result that
$\tau_{\alpha}^{LA}$ is smaller than $\tau_{\alpha}$ of depolarized
light-scattering spectra as shown in Fig.~1.  We then used the positions
of the $\alpha$ peaks in the MCT memory function to estimate the
relaxation time $\tau_{\alpha}^{MCT} = 1/\omega_{\alpha}^{MCT}$.
As seen in Fig.~1, the resulting $\tau_{\alpha}^{MCT}$ values 
follow the temperature dependence expected on the basis of the dielectric 
data of Schneider~{\it et al}.  We emphasize that this MCT analysis is the 
only one of the three employed in this study that can provide a meaningful 
estimate of $\tau_{\alpha}$.

The recent Brillouin scattering study of OTP by 
Monaco~{\it et al}~[Monaco 1999, Monaco 2000, Monaco 2001, Mossa 2001]
has suggested a different origin for the fast relaxation process observed
in Brillouin scattering spectra: coupling of longitudinal acoustic waves to 
intramolecular degrees of freedom as originally suggested by Mountain~[Mountain~1968].
Even in OTP, however, where this mechanism appears to be important,
some part of the fast relaxation must be associated with structural relaxation.  It is therefore an open question, that should be
carefully addressed, how much of the fast relaxation originates in each of
these two mechanisms, in OTP as well as in other materials -- including PC.

Finally, we return to the question of the ability of Brillouin scattering spectra
to extract meaningful information about the structural relaxation process.
In Fig.~11~ we show $\omega m''(\omega)$ (solid lines) and
$\omega m'(\omega)$ (broken lines) obtained from the MCT fits (top)
and the hybrid fits with $\beta = b$ (bottom).
%
\vbox{
\vspace{0.2 in}
\hbox{
\hspace{0in}
\epsfxsize 3.0in \epsfbox{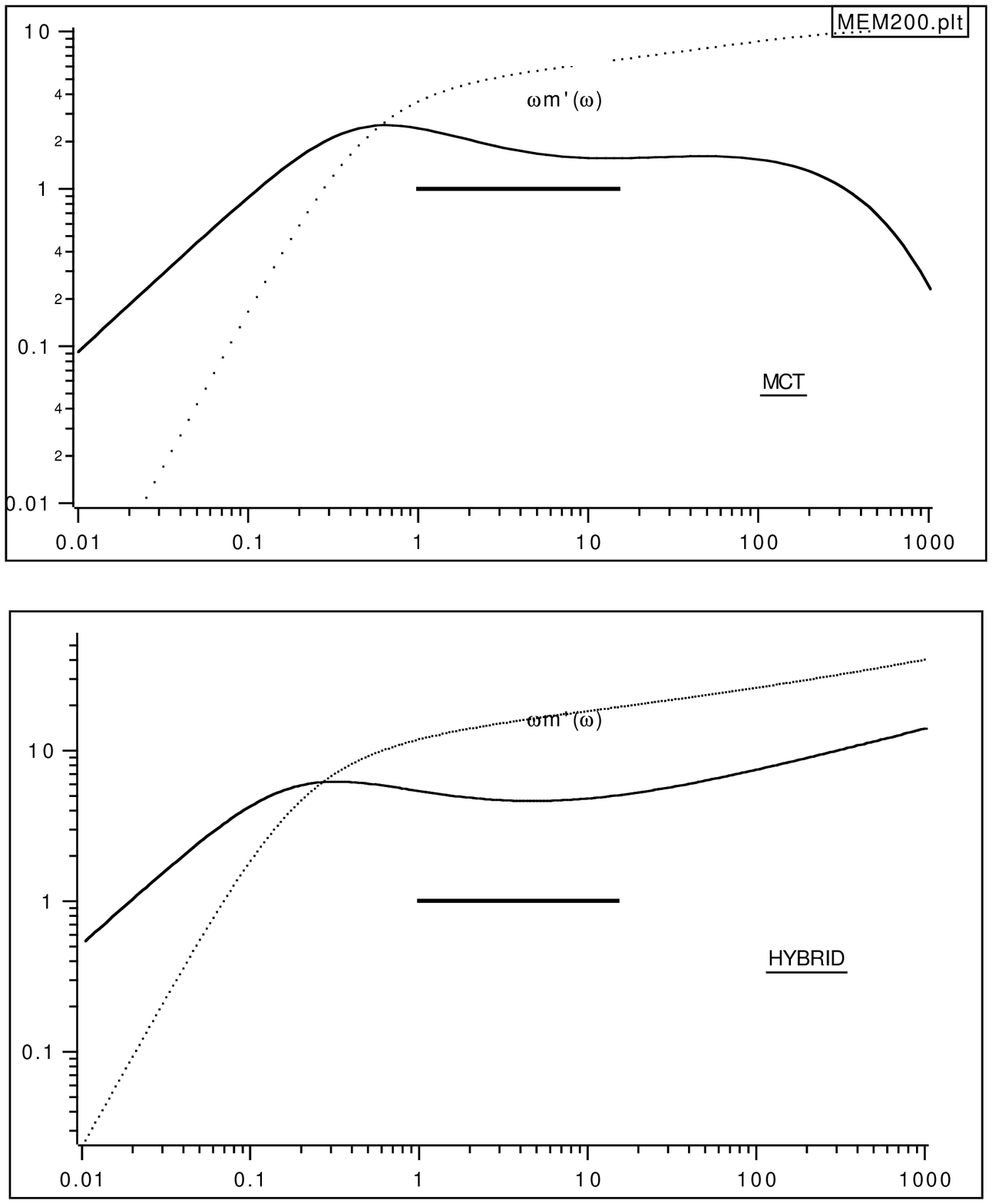}
}
\vspace{0.1in}
}
\refstepcounter{figure}
\parbox[b]{3.3in}{\baselineskip=12pt FIG.~\thefigure.
(MEMPLOT200).  $\omega m''(\omega)$ (solid lines) and
$\omega m'(\omega)$ (broken lines) from the fits at 200~K.
Top: MCT fit; bottom: hybrid fit with $\beta=b=0.5$ and 
$a=0.29$.  The fit range (1-15~GHz) is indicated in both plots 
by a thick horizontal line.}
\vspace{0.10in}

\label{1}
\vspace{0.1in}
  In the range of the Brillouin
spectra (1-15~GHz indicated by the horizontal solid lines), the shapes of 
these curves are remarkably close, even though they were obtained with 
very different procedures.  This result suggests that a reasonable
phenomenological model like the hybrid model, with the constraint that
$\tau_{\alpha}$ should have a temperature dependence consistent with 
that found by other techniques, can provide a credible representation of
the memory function $m(\omega)$ characterizing the longitudinal viscosity.

\acknowledgments

We thank W.~G\"{o}tze, J.~Wiedersich, M.~Fuchs, and
R. Pick for helpful discussions and suggestions, 
Th.~Voigtmann for providing the computer programs (and extensive advice)
for the schematic MCT analysis, and  P.~Lunkenheimer for providing
the $\tau_{\alpha}(T)$ values shown in Fig.~1.
HZC thanks the Alexander von Humboldt Foundation for support,
and the Technical University of Munich for hospitality during
a visit when this project was begun.

This research was supported by the National Science Foundation under 
Grants DMR-9616577 and DMR-9980370.
Travel support for the CCNY-TUM
collaboration was provided by NATO under Collaborative Research
Grant No.~CRG-930730. 

\section{Appendix:  Some Mathematical Details}

The normalized density fluctuation correlation function
\begin{equation}
\phi(t) = <\rho_{q}(0) \rho_{q}(t) >/<|\rho_{q}(0)|^{2} > 
 \ \ \ \ \ 
\end{equation}
obeys the generalized Langevin (or generalized oscillator) equation
\begin{equation}
\ddot{\phi}(t) + \gamma_{0}\dot{\phi}(t) + \omega_{0}^{2} \phi(t) +
\int^{t}_{0} m(t-t') \dot{\phi}(t') dt' = 0 
 \ \ \ \ \ \  
\end{equation}

We use the Laplace transform convention usually followed in the mode
coupling theory literature
\begin{equation}
F(z) = i \int^{\infty}_{0} e^{izt} F(t) dt 
\ \ \ \ \ \  
\end{equation}
with which we obtain
\begin{equation}
\phi(z) = \frac{-[z+m'(z)] - i[\gamma_{0}+m''(z)]}{[z^{2} -
 \omega_{0}^{2}+ z m'(z)] + i[z\gamma_{0}+z m''(z)]}
 \ \ \ \ \  
\end{equation}
where $m(z)=m'(z) +im''(z)$.

The power spectrum of the fluctuations in $\phi(t)$ is given by
\begin{equation}
S(w) = Im [\phi(z)]_{z = \omega} 
\ \ \ \ \ \  
\end{equation}
and the light scattering spectrum $I(\omega) \propto S(\omega)$ is then
\begin{equation}
I(\omega) = \frac{I_{0}[\gamma_{0}+m''(\omega)]}{[\omega^{2} -
\omega_{0}^{2} + \omega m'(\omega)]^{2} + [\omega \gamma_{0} + \omega m''
(\omega)]^{2}}  \ \ \ \ \ \   
\end{equation}
which is Eq.~(3) in the text.  
(Note that without the $i$ in Eq.~(18), $m'$ and
$m''$ would be exchanged, a convention that is frequently
employed in the Brillouin scattering literature.)

In the critical region preceding the $\alpha$ relaxation $m(t)$ can be represented by
\begin{equation}
m_{crit}(t) = t^{-a} e^{-t/\tau_{\alpha}}
\ \  \ \ \ \ 
\end{equation}
when the $e^{-t/\tau_{\alpha}}$ cutoff factor prevents $m_{crit}(t)$ from
adding to $m_{\alpha}(t)$ beyond the center of the $\alpha$ relaxation
region.  With Eq.~(18) and $m(\omega) = [m(z)]_{z=\omega}$ we have
\begin{equation}
m_{crit} (\omega) = i\int^{\infty}_{0} e^{i\omega t} t^{-a} e^{-t/\tau} dt 
 \ \ \ \ \ \ 
\end{equation}
from which [Gradshteyn~1965, p.~317] 
\begin{equation}
m_{crit}(\omega) = i \Gamma (1-a)
(\frac{1}{\tau} - i\omega)^{a-1}
 \ \ \ \ \ \  
\end{equation}
For the hybrid model, $m(\omega) = m_{CD}(\omega) +
 m_{crit}(\omega)$, so
\end{multicols}
\begin{eqnarray}
\omega m(\omega) & = & \Delta^{2} [ (1-i \omega \tau)^{-\beta} -1]
+ i \omega B \times\Gamma(1-a) [\tau^{-1} - i \omega]^{a-1}  \nonumber\\
 & = & \Delta^{2} [(1- i \omega \tau)^{-\beta} -1] + 
i \omega B \times \Gamma (1-a) \tau^{1-a} (1-i \omega \tau)^{a-1}
\ \ \ \ \ \ 
\end{eqnarray}
\begin{multicols}{2}
which is Eq.~(7) in the text.  (For $a = 0.29$, $\Gamma (1-a) = 1.282$.)

\newpage
\end{multicols}
\begin{table}
\caption{Properties of Propylene Carbonate}
\begin{tabular}{|l|c|c|}
Property & Value & Reference \\
\tableline
Formula & C$_{4}$H$_{6}$O$_{3}$ & \\
Molecular weight & 102.09 & \\
$T_{boil}$ & 513 K & \\
$T_{M}$ & 218 K & \\
$T_{G}$ & 160 K & \\
$T_{C}$ (MCT)  &  $185 \pm 5$ K  &  Gotze\tablenotemark[1] \\
Density $\rho (T)$ & $1.541 - 1.148 \times 10^{-3}$ T (K)(g/cm$^{3}$)
& Sim\tablenotemark[2] \\
Refractive index $n_{D}(T)$ & $1.5314-3.752 \times 10^{-4}$ T (K) & Sim  \\
Sound velocity $C_{0}(T)$ & ($2.5075 \times 10^{5} - 361.2 \times T)$~cm/sec
& Du\tablenotemark[3]  \\
$\beta_{K}$ (Kohlrausch stretching coefficient) & $0.77 \pm 0.05$ & Du   \\
$\beta_{CD}$ (Cole-Davidson stretching coefficient) &  0.68 & Du   \\
$T_{C}$ (MCT crossover temperature) & $187 \pm 5$~K & Du, Sch\tablenotemark[4] \\
  & 182 & Wut\tablenotemark[5]     \\
$\lambda$ (MCT exponent parameter) & $0.78 \pm 0.5$ & Du  \\
  & 0.72 & Wut  \\
$m$ (fragility index) & 104 & Bohm\tablenotemark[6] 

\tablenotetext[1]{(Gotze) W. G\"{o}tze and Th.~Volgtmann,
Phys.\ Rev.\ E {\bf 61}, 4133 (2000).}
\tablenotetext[2]{(Sim) L. Simeral and R.L. Amey, J. Phys.\ Chem.\ 
{\bf 74}, 1443 (1970).}
\tablenotetext[3]{(Du) W.M. Du, G. Li, H.Z. Cummins, M. Fuchs, J. Toulouse,
and L.A. Knauss, Phys.\ Rev.\ E {\bf 49}, 2192 (1994).}
\tablenotetext[4]{(Sch) U. Schneider, P. Lunkenheimer, R. Brand, and A. Loidl,
Phys.\ Rev.\ E {\bf 59}, 6924 (1999).}
\tablenotetext[5]{(Wut) J. Wuttke, M. Ohl, M. Goldhammer, S. Roth,
U. Schneider, R. Lunkenheimer, R. Kahn, B. Ruffl\'{e}, R. Lechner, and
M.A. Berg (submitted to Phys.\ Rev.\ E).}
\tablenotetext[6]{(Bohm) R. B\"{o}hmer, K.L. Ngai, C.A. Angell, and D.J. Plazek,
J.\ Chem.\ Phys.\ {\bf 99}, 4201 (1993).}
\end{tabular}
\end{table}

\tighten
\begin{table}
\caption{}
\begin{tabular}{|c||r|c|c|||c||r|r||r|r||c|c|r|c||} 
 $T(K)$ & \multicolumn{3}{c|||}{C.D. (free $\tau$)} &
\multicolumn{5}{c||}{Hybrid (fixed $\tau$)} &
\multicolumn{4}{c||}{Schematic MCT} \\ \cline{5-13} 
&\multicolumn{3}{c|||}{} & &
\multicolumn{2}{c||}{$\beta=0.68$} &
\multicolumn{2}{c||}{$\beta=b=0.5$} &
$\tau$ (ns) & $V_{s}^{DLS}$
& $V_{s}^{BR}$ & $\chi^{2}$  \\
\cline{1-13}\cline{1-13}

 & $\tau$\ (ns) & $\Delta^{2}$ (GHz$^{2}$) & $\chi^{2}$ & $\tau$ (ns) & $B/\Delta^{2}$ & $\chi^{2}$ & $B/\Delta^{2}$ & $\chi^{2}$ 
& & & & \\
\cline{1-13}\cline{1-13}
140 & 790 & 36.5 & 27.2 & $1.26 \times 10^{22}$ &  &  &  &  &  &  &  &  \\
\cline{1-13}
150 & 1220 & 40.2 & 27.8 & $1.42 \times 10^{14}$ & 0.01 & 27.4 & 0.002 & 28.6 &  & 94 &  &  \\
\cline{1-13}
160 & 31.8 & 37.0 & 22.1 & $2.49 \times 10^{8}$ & 0.04 & 19.5 & 0.034 & 22.7 &  & 80 & 30 & 25.2 \\
\cline{1-13}
165 & 14.2 & 34.3 & 23.3 & $1.96 \times 10^{6}$ & & & & & & & &  \\
\cline{1-13}
170  & 6.11 & 31.8 & 41.3 & $4.28 \times 10^{4}$  & 0.11 & 12.5 & 0.13 & 15.9 &  & 60 &  &  \\
\cline{1-13}
180  & 1.34 & 27.5 & 66.3 & $2.11 \times 10^{2}$  & 0.53 & 8.4 & 0.32 & 20.3 & $2.9 \times 10^{1}$ & 50 & 13 & 31.4 \\
\cline{1-13}
185  & 0.763 & 25.3 & 84.6 & $3.40 \times 10^{1}$  &  &  &  &  &  &  &  &  \\
\cline{1-13}
195  & 0.236 & 22.0 & 60.4 & 2.51 & 1.50 & 47.3 &  &  &  &  &  &  \\
\cline{1-13}
200  & 0.148 & 20.6 & 44.2 & $9.75 \times 10^{1}$  &  & 35.2 & 1.31 & 23.3 & $2.4 \times 10^{-1}$ & 30 & 7.6 & 5.85 \\
\cline{1-13}
205  & $1.03 \times 10^{-1}$ & 19.5 & 28.8 & $4.47 \times 10^{-1}$  &  &  &  &  &  &  &  &  \\
\cline{1-13}
210  & $7.28 \times 10^{-2}$  & 18.4 & 16.4 & $2.33 \times 10^{-1}$  & 4.00 & 10.1 & 2.72 & 8.3 &  & 25 & & \\
\cline{1-13}
215 & $5.41 \times 10^{-2}$ & 17.6 & 17.8 & $1.34 \times 10^{-1}$  &  &  &  &  &  &  &  &  \\
\cline{1-13}
220  & $4.15 \times 10^{-2}$  & 17.0 & 8.99 & $8.35 \times 10^{-2}$  &  & 19.2 & 18.1 & 5.75 & $3.0 \times 10^{-2}$ & 23 & 5.4 & 5.73 \\
\cline{1-13}
225 & $3.25 \times 10^{-2}$ & 16.9 & 9.43 & $5.55 \times 10^{-2}$ & 4.00 & 19.5 & & & & & & \\
\cline{1-13}
230  & $2.53 \times 10^{-2}$  & 16.7 & 5.21 & $3.89 \times 10^{-2}$  &  & 20.3 & 36.0 & 12.1 &  & 21 &  &  \\
\cline{1-13}
235  & $2.13 \times 10^{-2}$  & 17.0 & 5.35 & $2.85 \times 10^{-2}$  &  &  &  &  &  &  &  &  \\
\cline{1-13}
240  & $1.75 \times 10^{-2}$  & 17.0 & 11.9 & $2.16 \times 10^{-2}$  & 4.00 & 14.2 & 24.8 & 13.2 & $9.9 \times 10^{-3}$ & 18 & 4.7 & 11.9 \\
\cline{1-13}
250  & $1.28 \times 10^{-2}$  & 17.0 & 25.7 & $1.36 \times 10^{-2}$  &  &  &  &  &  & 16 &  &  \\
\cline{1-13}
260  & $9.96 \times 10^{-3}$  & 17.0 & 50.0 & $9.29 \times 10^{-3}$  & 4.00 & 41.6 & 21.2 & 41.3 & $4.7 \times 10^{-3}$ & 15 & 4.3 & 39.9 \\
\cline{1-13}
270  & $8.17 \times 10^{-3}$  & 17.0 & 47.4 & $6.80 \times 10^{-3}$  &  &  &  &  &  & 13 &  &  \\
\cline{1-13}
280  & $6.61 \times 10^{-3}$  & 17.0 & 66.3 & $5.24 \times 10^{-3}$  & 4.00 & 54.3 & 22.0 & 54.0 &  & 12 & $<3.5$ & 51.9 \\
\cline{1-13}
290  & $5.58 \times 10^{-3}$  & 17.0 & 60.5 & $4.19 \times 10^{-3}$  &  &  &  &  &  &  &  &  \\
\cline{1-13}
300  & $4.57 \times 10^{-3}$  & 17.0 & 60.7 & $3.46 \times 10^{-3}$  & 4.00 & 53.5 & 0.87 & 53.5 & & & $<2.5$ &  \\
\cline{1-13}
310  & $3.90 \times 10^{-3}$  & 17.0 & 43.4 & $2.93 \times 10^{-3}$  &  &  &  &  &  &  &  &  \\
\cline{1-13}
320  & $3.38 \times 10^{-3}$  & 17.0 & 49.2 & $2.53 \times 10^{-3}$  & 4.00 & 44.3 & 2.34 & 44.3 &  &  & $< 2.5$ & 40.5 \\
\cline{1-13}
330  & $2.76 \times 10^{-3}$  & 17.0 & 38.1 & $2.22 \times 10^{-3}$  &  &  &  &  &  &  &  &  \\
\cline{1-13}
340  & $2.26 \times 10^{-3}$  & 17.0 & 26.3 & $1.98 \times 10^{-3}$  &  &  &  &  &  &  &  &  \\
\cline{1-13}
350  & $1.80 \times 10^{-3}$  & 17.0 &  18.2 &
$1.79 \times 10^{-3}$  & 4.00 & 18.1 & 1.90 & 18.1 &  &  & $<2.5$ & 18 \\
\end{tabular}
\end{table}


\begin{references}
\bibitem{Note IIIB.1}[Note IIIB.1] In carrying out fits to
Eq.~(III.1) with the memory functions $m(\omega)$ of
Eqs.~(III.2) or (III.9), we used the following conventions for
units: $\omega$, $\omega_{0}$, $\gamma_{0}$, and $\Delta$
are all in GHz, while $\tau$ is in nsec.  In calculating 
$\omega \tau$, we therefore multiplied $\tau$ by $2\pi$.
\end{references}
\end{document}